# Evaluating E-OBS, AgERA5, MARS-STAT, ERA5, and ERA5-Land for Daily Minimum and Maximum Temperature Across Four Mediterranean European Countries

Dimitrios Voulanas

Texas A&M Energy Institute, Texas A&M University, College Station, TX, USA

* Correspondence: dvoulanas@tamu.edu

## Abstract

Gridded-dataset suitability varies by support, season, variable, and application. This study evaluated E-OBS, AgERA5, MARS-STAT/JRC Agri4Cast, ERA5, and ERA5-Land against observations from 624 European Climate Assessment & Dataset (ECA&D) and Hellenic National Meteorological Service (HNMS/EMY) stations, using 18.18 million source–station–day records containing minimum (TN) and maximum (TX) temperatures. Derived variables included mean temperature (Tmean), diurnal temperature range (DTR), and growing degree days (GDD). Identical-support rankings and station bootstraps assessed daily and GDD performance; Fisher's z transformation summarized seasonal correlations; temporal and network tests screened trends. E-OBS minimized daily and monthly-climatology-removed anomaly root mean square error (RMSE) for TN, TX, Tmean, and DTR in all 16 country-variable comparisons. With at least 10 matched years, E-OBS led 42 of 64 seasonal RMSE comparisons and 59 of 64 correlation comparisons; MARS-STAT/JRC Agri4Cast led the remaining 22 and five, respectively. E-OBS minimized GDD RMSE in Spain and Italy, whereas MARS-STAT/JRC Agri4Cast led in Greece and narrowly led in France. GDD RMSE equaled 3.3–7.2% of observed mean GDD. Robustness tests supported increases in Spanish annual DTR, French annual TX, and French growing-season GDD. The lapse-rate correction improved 22 of 40 comparisons but usually worsened TN. Dataset choice should reflect support, variable, season, terrain, and application.

Keywords: gridded temperature datasets; E-OBS; AgERA5; MARS-STAT; ERA5; ERA5-Land; diurnal temperature range; growing degree days; Mediterranean viticulture

## 1. Introduction

Daily minimum and maximum near-surface air temperatures (TN and TX, respectively) underpin climate monitoring, heat and cold exposure assessment, ecosystem analysis, agricultural planning, and the calculation of thermal indices. Station observations remain the local reference, but their distribution and continuity are uneven; gridded datasets therefore provide spatially complete fields and long records when station evidence alone is insufficient [1], [2]. The evaluated datasets represent distinct construction strategies. E-OBS interpolates European station observations and quantifies gridding uncertainty [2], [3]; ERA5 is a global atmospheric reanalysis, whereas ERA5-Land is a higher-resolution land-surface replay driven by ERA5 [4], [5]; and AgERA5 and MARS-STAT/JRC Agri4Cast provide agriculture-oriented temperature fields [6], [7]. Their source observations, model physics, station influence, temporal processing, and grid geometry differ sufficiently that analysts cannot treat the datasets as interchangeable.

The mismatch between a point observation and a grid-cell mean complicates evaluation against stations. Station density, changing network composition, interpolation smoothing, elevation, coastlines, islands, urban exposure, and land-sea representation can all alter apparent agreement. For E-OBS, sparse or changing station support increases gridding uncertainty and can smooth daily variability, extremes, and trends [8], [9], [10]. National high-resolution datasets over Iberia and Greece likewise show that good domain-average agreement can coexist with spatially heterogeneous errors in mountainous, coastal, and island environments [11], [12]. Reanalyses often reproduce the annual cycle and broad temporal variability, but regional biases persist over southern Europe and complex Mediterranean terrain [13]; finer nominal resolution alone does not remove errors associated with local topography, radiation, boundary-layer structure, or coastline placement [5], [14].

Fitness for use therefore depends on the application. Elevation correction can reduce large mountain-station errors, yet inversions and distinct local controls on daily minima and maxima can make fixed lapse rates fail [15], [16]. Derived quantities add another layer of sensitivity: DTR depends on the paired behavior of TN and TX, whereas GDD accumulates the daily mean-temperature excess above a stated base [17]. These distinctions are especially important for Mediterranean viticulture, where thermal zoning and growing-season interpretation depend on the temperature record and on contrasts among interior, coastal, island, and high-elevation areas [18], [19]. The four countries span Atlantic-influenced, continental, Mediterranean, alpine, coastal, and island environments, providing a stringent test of whether performance generalizes across contrasting viticultural regions.

A previous comparison of ERA-Interim, Agri4Cast, and E-OBS over Greek wine-producing regions found that relative performance changed with season, temporal scale, spatial characteristic, and derived drought index [20]. That precipitation-focused analysis established a useful multiscale evaluation sequence, but it did not determine the performance of current temperature datasets. A harmonized comparison of daily TN and TX, spatial fields, DTR, GDD, and trend behavior across several Mediterranean countries is still needed to distinguish general dataset behavior from country- and application-specific effects.

Accordingly, this study pursued two objectives. First, it compared the ability of E-OBS, AgERA5, MARS-STAT/JRC Agri4Cast, ERA5, and ERA5-Land to reproduce daily, monthly, seasonal, annual, and spatial TN and TX characteristics observed in Spain, France, Greece, and Italy. Second, it determined how temperature errors propagated into DTR and April–October GDD. The design preserved station-level traceability, used stations rather than daily rows as the sampling unit for uncertainty, and treated the uncorrected station-grid comparison as primary.

## 2. Materials and Methods

### 2.1 Study area and observed station network

The observed reference combined the ECA&D daily station archive for Spain, France, and Italy [1] with daily TN and TX observations that the Hellenic National Meteorological Service (HNMS; Greek acronym EMY) supplied for Greece. Stations entered the final network only after checks of coordinates, country assignment, variable availability, and temporal coverage. The final network contained 624 stations: 193 in Spain, 44 in France, 29 in Greece, and 358 in Italy. The source records spanned 1985–2018 in Spain, 1985–2012 in France, 1981–2004 for Greek TN, 1981–2012 for Greek TX, and 1985–2015 in Italy.

Table 2 records source availability separately from effective paired support. It distinguishes the longer Greek TX archive (through 2012) from the paired TN–TX analytical period (through 2004). Spain, France, and Italy use ECA&D observations; Greece uses HNMS/EMY observations.

The station network spans continental interiors, mountain environments, long coastlines, peninsulas, and islands. Station elevations ranged from −4 to 2,535 m, and the number of eligible stations differed substantially among countries. These contrasts support evaluation across different combinations of terrain, station density, and land-sea geometry. The spatial interpolation omitted twelve Spanish stations outside the mainland-and-Balearic plotting domain; the temporal analyses retained their records.

### 2.2 Temperature datasets

This study evaluated five datasets representing three construction strategies. E-OBS supplied station-interpolated daily TN and TX [2], [3]. AgERA5 supplied local-time daily statistics from a production system that used hourly ERA5 fields, interpolated them to a 0.1° grid, and adjusted them using grid- and variable-specific relationships calibrated against high-resolution forecasts from the European Centre for Medium-Range Weather Forecasts (ECMWF) [6]. MARS-STAT/JRC Agri4Cast supplied daily station-interpolated European agrometeorological fields on a 25 × 25 km grid [7]. ERA5 and ERA5-Land supplied hourly 2 m air temperature fields, which preprocessing scripts aggregated to daily TN and TX [4], [5]. Table 1 summarizes dataset characteristics and station matching.

Temperature values use degrees Celsius, and the archive records dataset, country, station, date, and source version. ERA5 and ERA5-Land daily extrema retain the calendar-date labels stored in the archive. Missing hourly timestamps, provider time-zone metadata, and station observation-day conventions prevented reaggregation and a sensitivity test of Coordinated Universal Time (UTC) versus local-standard-time calendar days. The Supplementary Methods document this limitation.

### 2.3 Observed-data quality control, same-date pairing, and completeness

All calculations began with TN/TX pairs matched by station and date. A daily record entered the analysis only when observed and gridded values of TN and TX were available for the same country, station, and date; the archive retained the dataset source version as metadata. Quality control excluded 205 Italian ECA&D station-days for which TN exceeded TX and screened for duplicate keys and non-finite or physically implausible values. The quality-control audit identified the TN > TX inconsistency but not its cause, so the study made no more specific attribution. Because the archive lacks provider-level quality-control flags and station relocation metadata, Pettitt change-point tests with false discovery rate control screened annual station series for potential breakpoints. This screen identified potential breaks but did not replace formal homogenization.

A month qualified as valid when it contained at least 80% of the expected paired daily values. Seasonal and annual values required every constituent month to be valid. The study divided the April–October growing season into spring (April–May), summer (June–August), and autumn (September–October), following the analytical periods used in the earlier Greek gridded-dataset comparison [20]. These fixed calendar subdivisions represented viticulture-oriented thermal periods rather than direct observations of grapevine phenological stages.

Two support modes served distinct purposes. Maximum-coverage analyses described each dataset's available support, whereas direct rankings used the station-date intersection shared by the observed reference and all five datasets. This common-support rule governed daily, daily-anomaly, seasonal, GDD, Winkler-region, frost-day, and hot-day comparisons; paired station resampling quantified uncertainty and first-rank probabilities. Although Greek TX source availability extended to 2012, every paired TN–TX analysis ended in 2004 with the corresponding TN record.

## 2.4 Station-to-grid matching

The station-matching procedure linked each station to its corresponding grid cell or to a supplied station-coordinate extract. The mapping retained station and dataset identifiers, grid coordinates, elevation, and separation, and the same mapping supported all analyses. The MARS-STAT/JRC Agri4Cast extract supplied values at station coordinates but did not retain native grid-cell centers. Because those centers were unavailable, physical station-grid distances could not be reconstructed and were reported as unavailable rather than 0 km.

A reproducible station-grid mapping does not make point and grid-cell measurements physically equivalent. A station represents its immediate exposure, whereas a grid cell averages terrain, land cover, and atmospheric conditions over a finite area. The resulting validation metrics reflect both dataset error and point-grid representativeness; they do not isolate either component. Mapping distance served as a diagnostic, not a correction or a sufficient explanation for local disagreement.

E-OBS and MARS-STAT/JRC Agri4Cast are partly station-derived. Because the archive lacks exact construction-station membership, it cannot separate independent from in-network evaluation. Accordingly, the study reports agreement with the available station reference rather than fully independent validation.

## 2.5 Daily, monthly, seasonal, and annual aggregation and derived temperature indices

For every retained daily pair, Equation (1) defines mean temperature as

$$\mathrm{Tmean}_d = \frac{\mathrm{TN}_d + \mathrm{TX}_d}{2} \tag{1}$$

Equation (2) defines daily diurnal temperature range as

$$\mathrm{DTR}_d = \mathrm{TX}_d - \mathrm{TN}_d \tag{2}$$

Equation (3) accumulates growing degree days over April–October using a 10 °C base:

$$\mathrm{GDD}_y = \sum_{d \in \mathrm{Apr-Oct}} \max(\mathrm{Tmean}_d - 10, 0) \tag{3}$$

The daily formulation in Equation (3) preserves threshold crossings that monthly or seasonal mean-temperature calculations would lose [17].

For an application-oriented sensitivity analysis, the study assigned each station-year GDD value to one of the five classical Winkler regions using boundaries at 1390, 1670, 1940, and 2220 degree-days [18]. For each station, same-region agreement equaled the proportion of matched years in which the dataset and observed reference occupied the same region; country summaries then averaged these station-level proportions with equal weight. Supplementary diagnostics counted April–May frost days (TN < 0 °C) and June–August hot days (TX ≥ 35 °C) on identical support. These operational thresholds do not represent cultivar-specific phenology.

Aggregation followed daily pairing. Monthly summaries averaged valid daily TN, TX, Tmean, and DTR values; seasonal and annual summaries required every constituent month to be valid. Country-year series assigned equal weight to each station so that long or highly complete records did not dominate national means. For each station-year, daily positive temperature excesses above the 10 °C base were summed to obtain growing-season GDD before country-level averaging.

### 2.6 Elevation correction

To isolate one component of point-grid mismatch, the study compared station elevation with the elevation at each matched dataset location. For ERA5, the matching procedure extracted surface geopotential from the temperature grid cell and converted it to height by dividing by standard gravity (9.80665 m $s^{-2}$) following ECMWF guidance [21]. The matching procedure yielded grid and station elevations for all 3,120 station-dataset combinations. The fixed sensitivity adjustment was

$$T_{\text{corr}} = T_{\text{raw}} + 0.0064\left(z_{\text{grid}} - z_{\text{station}}\right) \tag{4}$$

where elevations are in meters and temperature is in degrees Celsius. Observed station temperatures remained unchanged. Because environmental lapse rates vary with season, location, atmospheric stability, and temperature variable, the study evaluated Equation (4) only as a sensitivity test; the uncorrected comparison remained primary [15], [16].

### 2.7 Validation metrics and Taylor diagnostics

For each station, date, dataset, and variable, the study defined error as the dataset value minus the observed value. Equations (5) and (6) define bias and RMSE:

$$\text{Bias} = \frac{1}{n}\sum_{i=1}^{n}(P_i - O_i) \tag{5}$$

$$\text{RMSE} = \sqrt{\frac{1}{n}\sum_{i=1}^{n}(P_i - O_i)^2} \tag{6}$$

Here, $P_i$ is the dataset value, $O_i$ is the observed value, and n is the number of matched values. The validation also included mean absolute error (MAE), centered RMSE, Pearson correlation, and the dataset-to-observed standard-deviation ratio. Direct daily comparisons used the intersection of all five datasets and the observed reference, first at each station and then with equal station weight. Paired station bootstraps provided 95% intervals and the probability that each dataset ranked first. Daily anomaly validation subtracted each station-dataset monthly climatology before recalculating the metrics, thereby separating weather-scale departures from the shared annual cycle.

Seasonal validation used matched station-year-period evidence common to the observed reference and all five datasets. The primary analysis required at least 10 matched years per station, and a five-year threshold provided sensitivity evidence. Seasonal validation transformed station correlations with Fisher's z, averaged the transformed values with equal station weight, and back-transformed the mean to r. The median standard-deviation ratio and median absolute natural log-ratio summarized variability; exact agreement corresponds to a ratio of 1 and an absolute log-ratio of 0. Paired station bootstraps supplied intervals and rank probabilities. Because the primary 10-year filter reduced Greek seasonal support to four stations, the interpretation also considered the five-year sensitivity results.

Country-level Taylor diagrams combined Pearson correlation, normalized standard deviation, and normalized centered RMS difference. The observed reference lies at correlation 1 and normalized standard deviation 1; the radial coordinate represents the dataset-to-observed standard-deviation ratio, the angular coordinate represents correlation, and dashed contours show normalized centered RMS difference. The diagrams provide a compact synthesis of temporal coherence and amplitude but do not replace bias or RMSE.

### 2.8 Trend analysis

Equal-weight station averages formed annual and seasonal country-year series, and ordinary least-squares (OLS) regressions estimated slopes per decade. Inference combined Newey–West heteroskedasticity- and autocorrelation-consistent (HAC) intervals, station-bootstrap intervals, Benjamini–Hochberg false discovery rate (FDR) adjustment, and a fixed-network series. The fixed network required valid values in at least 80% of years and at least one valid value in both the first and last 20% of the country record. The Benjamini–Hochberg procedure controlled FDR across the 80 annual and seasonal temperature tests and, separately, across the four country-level GDD tests. Representativeness checks used one-degree spatial-cell weighting, elevation-tertile estimates, and the Spanish mainland-and-Balearic map domain. A change was classified as directionally robust when the HAC test remained significant after FDR control, the station-bootstrap interval excluded zero, and the fixed-network slope retained the same sign. The study reported fixed-network significance separately and excluded it from the directional criterion.

Station-specific regressions described local heterogeneity in annual TN, TX, Tmean, DTR, and April–October GDD when at least 10 valid annual values were available. Supplementary Figures S1–S5 and S15–S17 show local patterns, record-length sensitivity, and all trend-error observations. These regressions remain descriptive because record windows differ and the Pettitt screen identified potential inhomogeneities at a subset of stations.

### 2.9 Spatial interpolation

Inverse-distance weighting (IDW) generated separate spatial fields for each country and variable. Within each six-panel country figure, all panels used a common station set, geographic domain, temperature scale, station markers, and dataset-minus-observed contours. IDW provided a deterministic distance-weighted summary for irregularly spaced observations [22], [23]. Because an alternative radial-basis-function interpolation produced values outside the range spanned by the input station values, the study excluded it. The resulting maps are descriptive visualizations of the station evidence; they do not substitute for climate datasets or independent validation fields.

***Table 1. Evaluated temperature datasets and station-grid mapping. *MARS-STAT/JRC Agri4Cast was supplied at station coordinates; its native-grid separation was not recoverable.***

| Dataset | Dataset family | Temporal basis | Station-grid mappings | Mean grid distance (km) |
|---|---|---|---|---|
| AgERA5 | ERA5-derived agroclimatic dataset adjusted against ECMWF high-resolution forecasts | daily local-time TN/TX on a 0.1° grid | 624 | 3.646 |
| E-OBS | station-interpolated gridded observations | daily TN/TX | 624 | 3.755 |
| ERA5 | global atmospheric reanalysis | hourly 2 m temperature aggregated to daily TN/TX | 624 | 9.308 |
| ERA5-Land | ERA5-driven land-surface reanalysis | hourly 2 m temperature aggregated to daily TN/TX | 624 | 3.646 |
| MARS-STAT/JRC Agri4Cast | station-interpolated agrometeorological dataset | daily TN/TX on a 25 × 25 km grid; analytical extract at station coordinates | 624 | N/A* |
| Observed stations | station observations (ECA&D; HNMS/EMY in Greece) | daily TN/TX station observations | 624 | — |

***Table 2. Observed station coverage, source availability, paired TN–TX analytical period, and elevation range. Eligible denotes stations with at least one valid paired record; Spatial denotes stations included in the plotted country domain.***

| Country | Eligible | Spatial | Source TN availability | Source TX availability | Paired analytical period | Min elev. (m) | Median elev. (m) | Max elev. (m) |
|---|---|---|---|---|---|---|---|---|
| Spain | 193 | 181 | 1985–2018 | 1985–2018 | 1985–2018 | 1 | 336 | 2535 |
| France | 44 | 44 | 1985–2012 | 1985–2012 | 1985–2012 | 2 | 101.5 | 871 |
| Greece | 29 | 29 | 1981–2004 | 1981–2012 | 1981–2004 | 0 | 32 | 662 |
| Italy | 358 | 358 | 1985–2015 | 1985–2015 | 1985–2015 | −4 | 149 | 2165 |

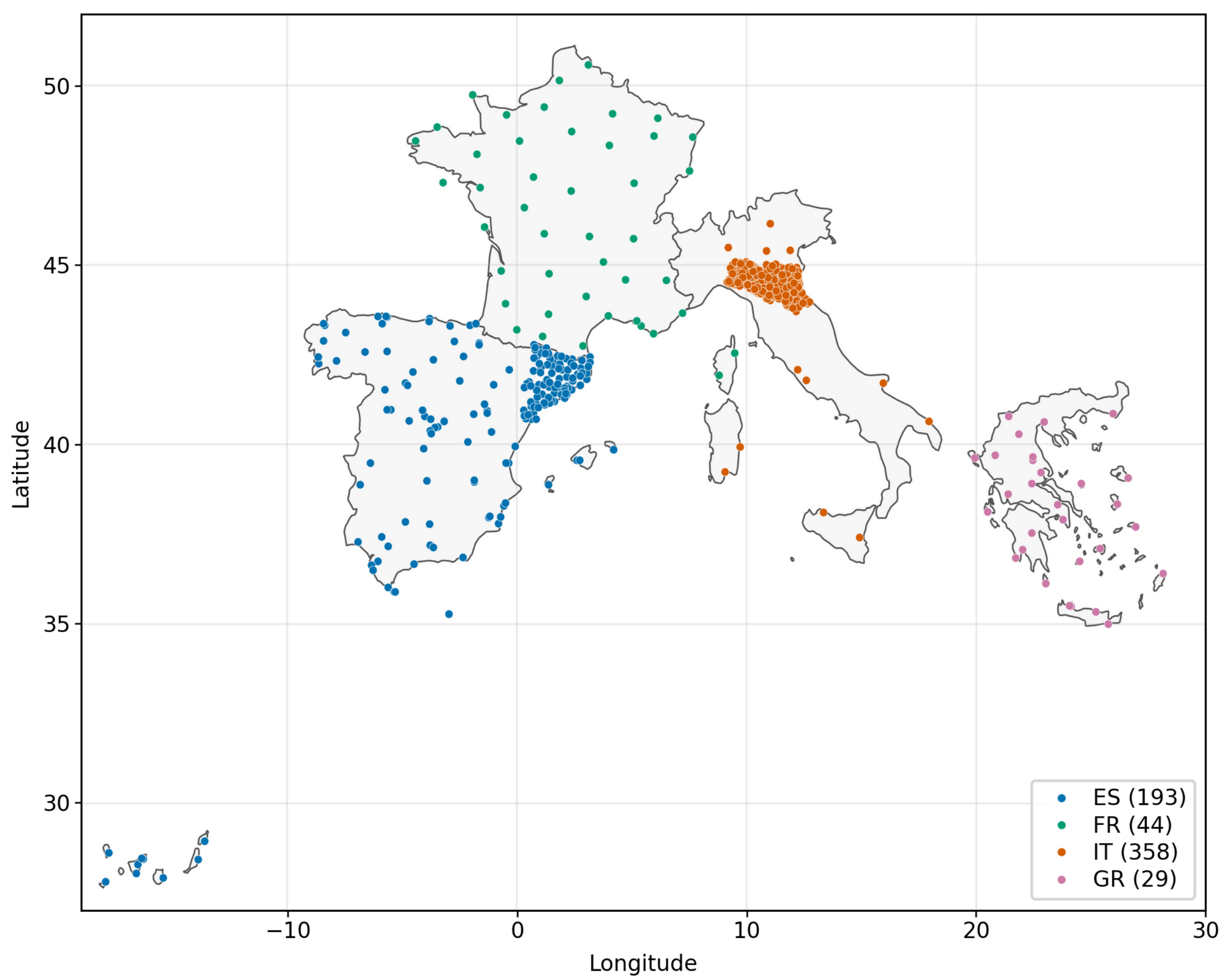

*Figure 1. Study domain and the 624 observed stations. Spain, France, and Italy use ECA&D observations; Greece uses HNMS/EMY observations. ES, FR, GR, and IT denote Spain, France, Greece, and Italy, respectively.*

# 3. Results

The validated analytical archive contains 18,179,949 source–station–day records with paired TN and TX values: 3,148,051 observed-reference records and 15,031,898 gridded-dataset records. This total does not represent statistically independent observations. Identical daily support across all five datasets retained 159 stations in Spain, 37 in France, 13 in Greece, and 351 in Italy. After requiring at least five valid growing seasons per station, the common-support GDD analysis retained 144, 36, 10, and 261 stations, respectively. Daily, daily-anomaly, seasonal-mean, GDD, Winkler-region, frost-day, and hot-day rankings used identical cross-dataset support. Monthly climatology, trend, and spatial analyses used the analysis-specific support described in the Methods and the relevant captions.

## 3.1 Annual and viticulture-oriented seasonal trends

Observed annual slopes were heterogeneous, but robustness tests reduced the set of defensible findings (Table 3; Figure 2; Supplementary Figure S13). Spanish DTR increased by 0.163 °C decade$^{-1}$, and French TX increased by 0.360 °C decade$^{-1}$ under the predefined directional-robustness criterion: HAC tests remained significant after FDR control, station-bootstrap intervals excluded zero, and fixed-network slopes retained the same sign. For both changes, the corresponding fixed-network HAC tests were also significant. French TN and Tmean remained positive but failed at least one criterion. Spanish TN and Tmean declines and Italian TN and DTR changes reversed or weakened under fixed-network or spatial-weighting sensitivity, while all Greek annual changes remained inference-sensitive.

During the April–October growing season, directionally robust seasonal signals occurred in specific periods rather than throughout the growing season (Supplementary Figures S6, S7, and S17). France retained increases in spring TN, TX, Tmean, and DTR, as well as in growing-season TN, TX, and Tmean; the corresponding fixed-network HAC tests were also significant. Greek summer Tmean increased by 0.481 °C decade$^{-1}$ and met the directional criterion: the HAC test remained significant after FDR control, the station-bootstrap interval excluded zero, and the fixed-network slope retained the same sign. Separate Greek TN and TX changes were network-sensitive. Spain retained a directionally robust spring DTR increase, although its fixed-network HAC test was not significant. Italy's growing-season, spring, and summer DTR slopes retained positive signs, but the fixed-network slopes were small and nonsignificant.

Large OLS point estimates that failed the full robustness screen remain descriptive rather than conclusive. Examples include Italian spring and summer TX warming, autumn TN cooling, and several Greek seasonal TN/TX changes. Network-composition diagnostics (Supplementary Figure S12) clarify this instability: fixed-network and one-degree-cell-weighted slopes often differed from the all-available-station estimate, especially in Spain and Italy.

Because many observed target slopes were themselves network- or inference-sensitive, gridded-dataset slope rankings remained secondary. Relative to the all-available-station OLS targets, ERA5 and ERA5-Land often produced the closest slopes, but no dataset led consistently across directionally robust changes, variables, periods, and countries. Trend studies should therefore evaluate the observed reference, matched support, and dataset slope together rather than rank datasets against a single unstable target.

***Table 3. Observed annual trends. n denotes stations contributing eligible annual series to the country-year analysis. "Directionally robust" requires HAC significance after FDR control, a station-bootstrap interval excluding zero, and fixed-network sign agreement; fixed-network significance is reported separately and is not required. Slopes and intervals are °C decade$^{-1}$. "Sensitive" denotes failure of at least one directional-robustness criterion or sign instability under the fixed-network or spatial-weighting checks.***

| Country | Metric | n | OLS slope | HAC 95% CI | Station-bootstrap 95% CI | Assessment |
|---|---|---|---|---|---|---|

| Country | Metric | n | OLS slope | HAC 95% CI | Station-bootstrap 95% CI | Assessment |
|---|---|---|---|---|---|---|
| Spain | TN | 187 | −0.355 | −0.595 to −0.114 | −0.592 to −0.159 | Sensitive |
| Spain | TX | 187 | −0.191 | −0.410 to 0.027 | −0.440 to 0.053 | Sensitive |
| Spain | Tmean | 187 | −0.273 | −0.500 to −0.046 | −0.504 to −0.071 | Sensitive |
| Spain | DTR | 187 | +0.163 | 0.095 to 0.232 | 0.034 to 0.324 | Directionally robust |
| France | TN | 43 | +0.266 | −0.010 to 0.542 | 0.149 to 0.356 | Sensitive |
| France | TX | 43 | +0.360 | 0.069 to 0.651 | 0.247 to 0.453 | Directionally robust |
| France | Tmean | 43 | +0.313 | 0.035 to 0.590 | 0.205 to 0.398 | Sensitive |
| France | DTR | 43 | +0.094 | −0.024 to 0.212 | 0.038 to 0.160 | Sensitive |
| Greece | TN | 29 | +0.077 | −0.157 to 0.311 | −0.791 to 0.698 | Sensitive |
| Greece | TX | 29 | −0.054 | −0.278 to 0.170 | −0.292 to 0.258 | Sensitive |
| Greece | Tmean | 29 | +0.011 | −0.211 to 0.234 | −0.488 to 0.449 | Sensitive |
| Greece | DTR | 29 | −0.131 | −0.240 to −0.023 | −0.766 to 0.643 | Sensitive |
| Italy | TN | 320 | −0.464 | −0.903 to −0.025 | −1.217 to 0.590 | Sensitive |
| Italy | TX | 320 | +0.098 | −0.316 to 0.513 | −0.876 to 1.352 | Sensitive |
| Italy | Tmean | 320 | −0.183 | −0.587 to 0.222 | −1.060 to 1.016 | Sensitive |
| Italy | DTR | 320 | +0.562 | 0.287 to 0.837 | 0.089 to 1.007 | Sensitive |

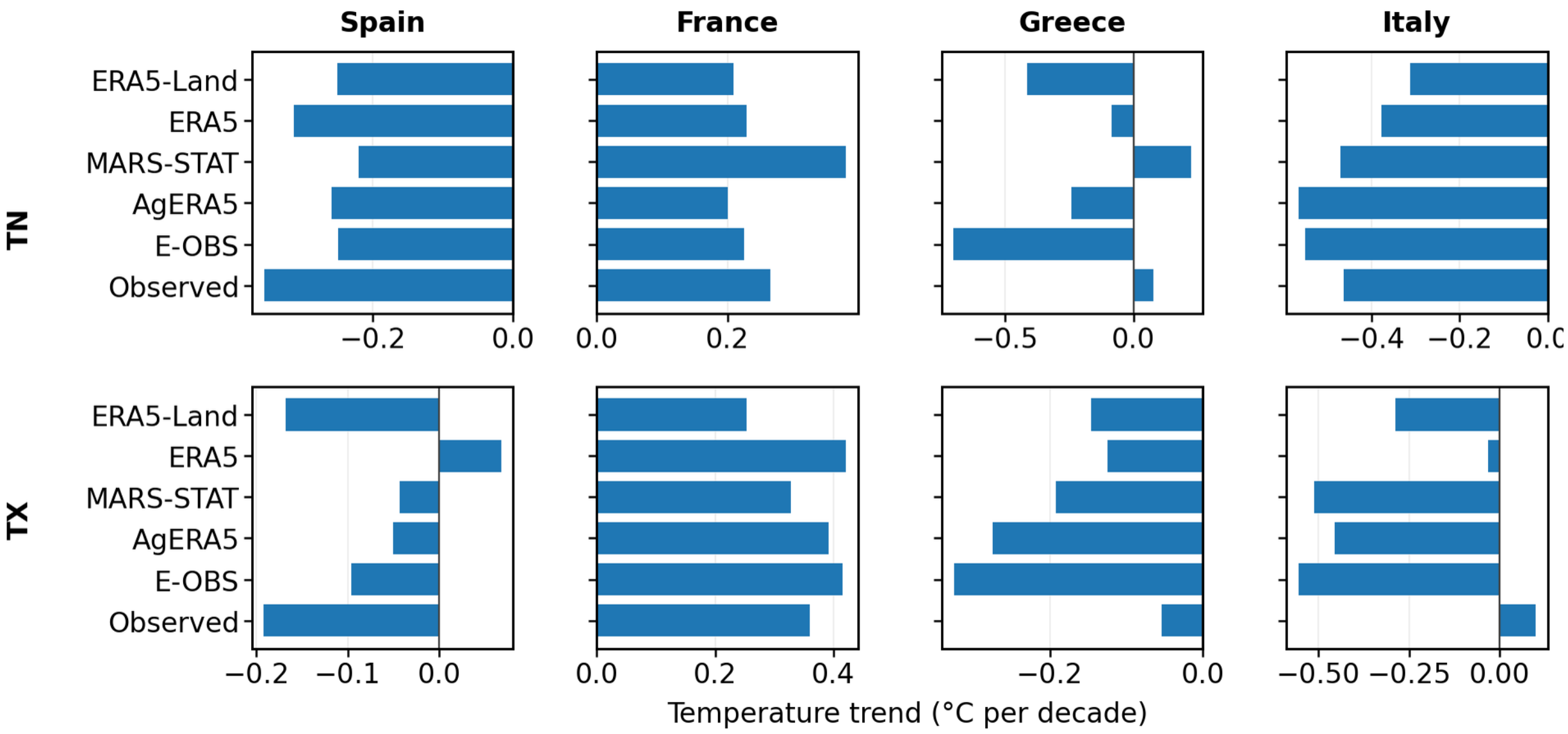

*Figure 2. Annual TN and TX trends from station-balanced country-year means. Table 3 reports observed annual slopes; Supplementary Figures S6–S7 and S13 show the corresponding seasonal point estimates and trend-sensitivity diagnostics. MARS-STAT denotes MARS-STAT/JRC Agri4Cast.*

## 3.2 Monthly climatology and seasonal validation performance

Using each dataset's maximum eligible paired station support, all datasets reproduced the annual cycle of TN and TX, but monthly offsets differed by country (Figure 3). MARS-STAT/JRC Agri4Cast most closely followed the French cycle, with mean dataset-minus-observed TN and TX differences of +0.080 and −0.008 °C; its corresponding Greek differences were also small (+0.194 and −0.295 °C). E-OBS was nearly unbiased in Italy (TN, −0.030 °C; TX, approximately 0 °C) and showed a small mean difference for Spanish TX (−0.035 °C), but its Greek TN and TX climatologies were cooler by 1.83 and 0.86 °C. Thus, low daily RMSE did not always imply an unbiased monthly cycle.

The three ERA5-derived datasets showed a consistent cool bias in monthly TX. Country-average monthly TX differences ranged from −2.16 to −0.29 °C, with the largest departures in Greece and the smallest in Italy. TN offsets were less uniform. The preferential TX underestimation narrowed the modeled seasonal TN–TX separation and anticipated the negative DTR biases in the daily analysis.

Applying the 10-year threshold changed seasonal rankings (Supplementary Figures S8–S11). E-OBS led 42 of 64 RMSE comparisons, 59 of 64 correlation comparisons, and 50 of 64 variability comparisons based on the median absolute log-ratio; MARS-STAT/JRC Agri4Cast led the remaining 22, five, and 14. Under the five-year threshold, the RMSE split was 47 to 17.

E-OBS led all 16 RMSE comparisons in Spain and Italy. MARS-STAT/JRC Agri4Cast led 12 of 16 RMSE comparisons in France and 10 of 16 in the primary Greek subset of four stations; E-OBS led the remainder. In France, MARS-STAT/JRC Agri4Cast led TX, Tmean, and DTR for most periods. In Greece, rankings changed by variable and between the five- and 10-year thresholds. These sensitivities preclude treating the Greek seasonal ranking as a stable national hierarchy.

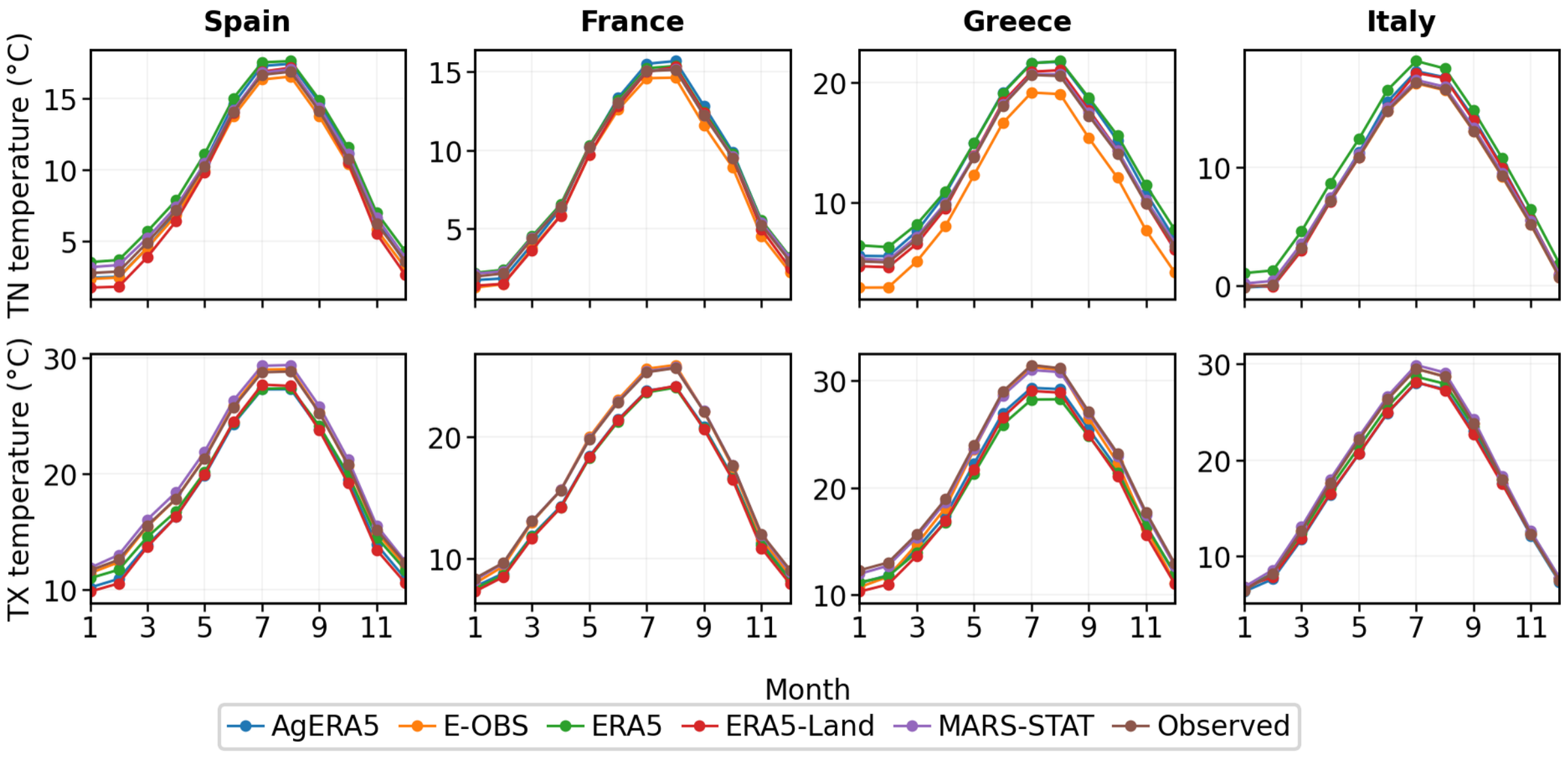

***Figure 3. Monthly station-balanced TN and TX climatology by country, calculated using each dataset's maximum eligible paired station support. Dataset-specific station support therefore differs, so the figure is descriptive rather than a direct common-support ranking. Greek observations are from HNMS/EMY; the other countries use ECA&D. MARS-STAT denotes MARS-STAT/JRC Agri4Cast.***

## 3.3 Local station-scale trend heterogeneity

Station-specific annual slopes revealed local warming signals that country means did not consistently capture (Supplementary Figures S1–S3 and S15–S16). Median local Tmean slopes were positive in all countries, but record length and potential breakpoints differed substantially. The separate TN and TX maps show that local changes in minima and maxima followed different spatial patterns, particularly in Italy and Greece.

Local DTR and seasonal slopes were more mixed (Supplementary Figures S2 and S17). Positive and negative station trends coexisted within each country, and station-slope distributions remained widest where network composition changed most, even after requiring at least 10 valid years. These patterns explain why country-level DTR and seasonal slopes can be sensitive even when many local slopes share one sign.

Local GDD slopes were predominantly positive, but only French country-level GDD warming met the HAC, station-bootstrap, and fixed-network criteria (Supplementary Figure S3; Table 5). Spain, Greece, and Italy had network-sensitive country trends despite positive local medians, demonstrating that unequal record windows and changing station support can reverse aggregate signs.

E-OBS had the smallest local Tmean and GDD slope MAE in Spain, France, and Italy, whereas MARS-STAT/JRC Agri4Cast had the lowest MAE in Greece (Supplementary Figures S4 and S5). Supplementary Figures S4 and S5 display all retained station points, including values beyond conventional boxplot whiskers.

### 3.4 Spatial temperature patterns

Growing-season spatial fields reproduced the dominant observed gradients but differed around mountain belts, coastlines, and islands (Figures 4–11). Spain was warmer in the south and southeast and cooler in the northern interior and mountains; France showed a north-south contrast with warmer Mediterranean and Corsican conditions; Greece showed cooler northern and upland mainland conditions and warmer southern and island environments; and the Italian fields resolved strong Alpine and Apennine gradients. E-OBS generally followed these patterns closely, while the reanalysis-derived datasets showed broader departures, especially for TX. Because the maps interpolate station evidence rather than independent spatial observations, they are descriptive spatial summaries rather than validation fields.

Spain contributed 181 stations to the mapped domain, with observed elevations reaching 2,535 m. The principal departures occurred across high terrain and transition zones between the cool northern interior and warmer Mediterranean south and southeast. TX differences were more spatially coherent than TN differences, particularly for the ERA5-derived datasets (Figures 4 and 8).

France contributed 44 stations spanning elevations from 2 to 871 m. E-OBS and MARS-STAT/JRC Agri4Cast retained the cooler northern and western conditions and the warmer Mediterranean and Corsican sectors, whereas broader TX departures appeared in the southeast, interior relief, and Corsica (Figures 5 and 9).

Greece combined only 29 HNMS/EMY stations with complex mainland relief, coastlines, and islands. The network sampled elevations of 0–662 m and therefore did not validate the country's highest mountain environments. Within the sampled range, all datasets reproduced the cooler north–warmer south contrast, but departures differed over northern relief, the Peloponnese, Crete, and the island domain (Figures 6 and 10).

Italy had 358 stations and elevations up to 2,165 m. The fields resolved the Alps, Apennines, warmer south, Sicily, and Sardinia; cool TX departures in the ERA5-derived datasets were consistent with Italy's larger TX errors relative to France (Figures 7 and 11).

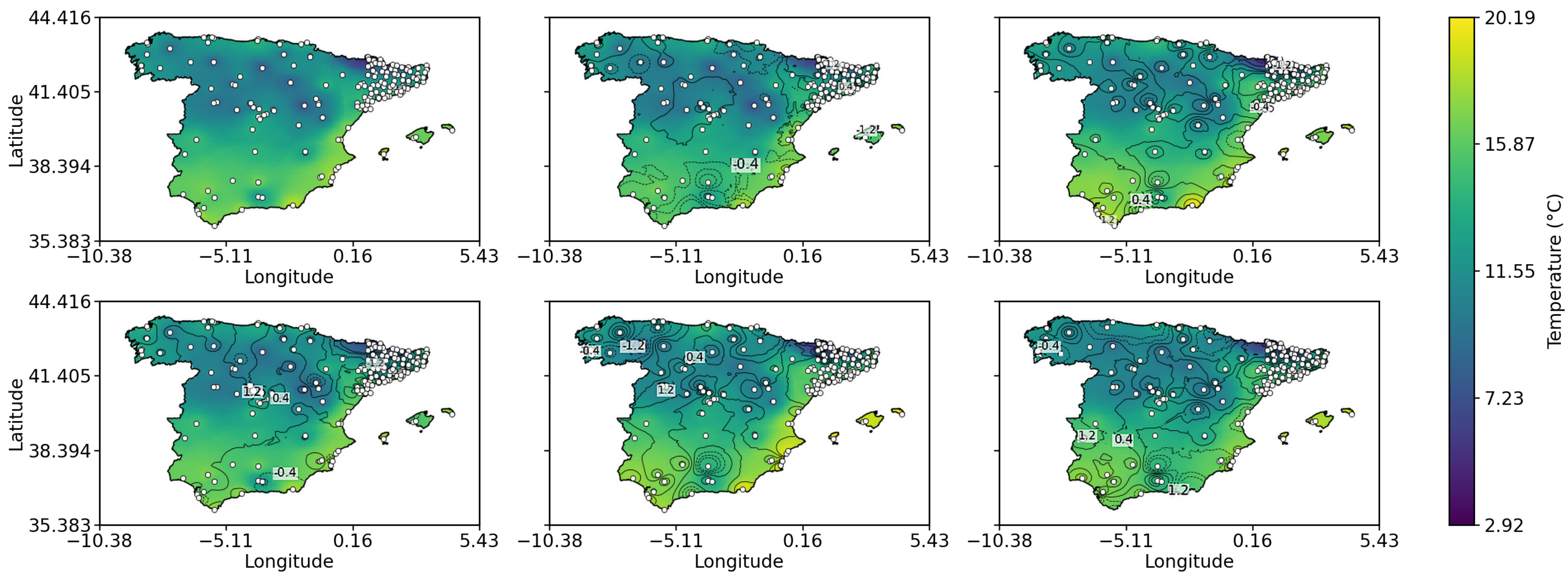

*Figure 4. Spain growing-season uncorrected TN spatial fields using IDW. The observed panel uses ECA&D stations. The panels, read from left to right and top to bottom, show Observed, E-OBS, AgERA5, MARS-STAT/JRC Agri4Cast, ERA5, and ERA5-Land; contours show dataset-minus-observed differences; all temperatures are in °C.*

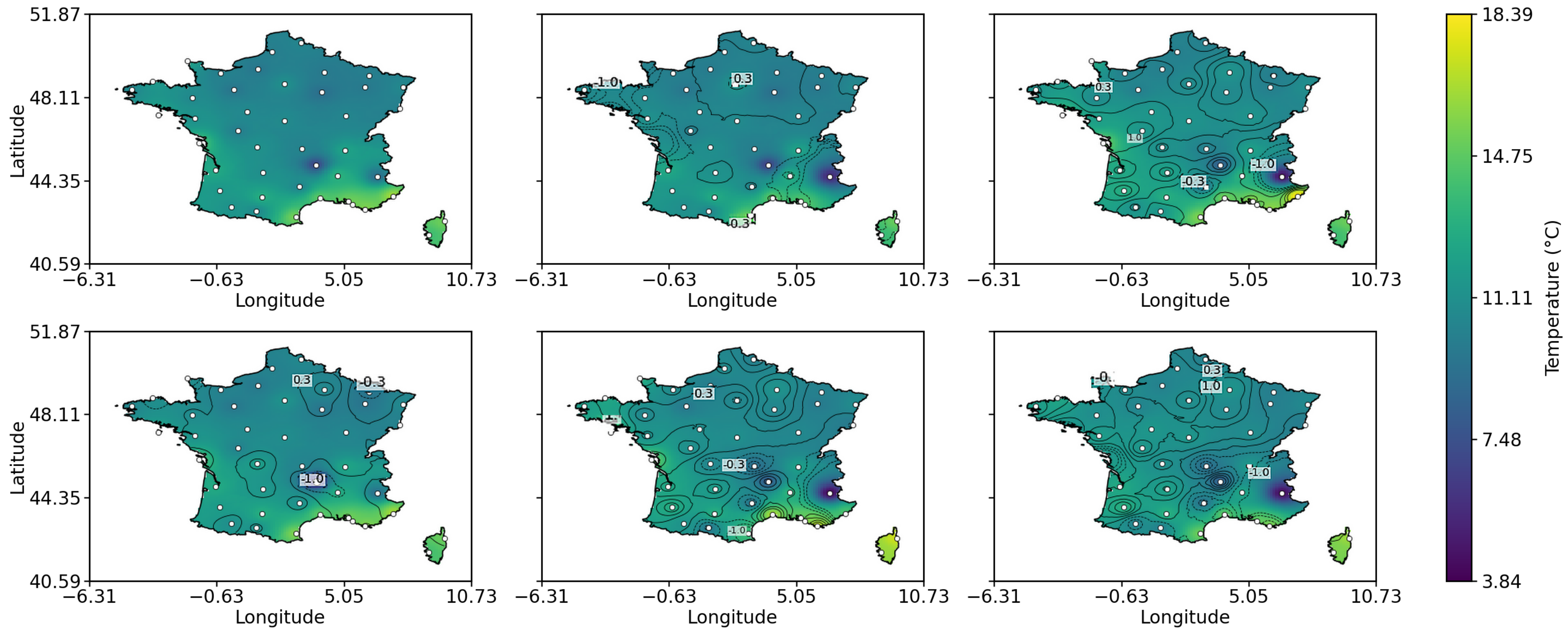

***Figure 5. France growing-season uncorrected TN spatial fields using IDW. The observed panel uses ECA&D stations. The panels, read from left to right and top to bottom, show Observed, E-OBS, AgERA5, MARS-STAT/JRC Agri4Cast, ERA5, and ERA5-Land; contours show dataset-minus-observed differences; all temperatures are in °C.***

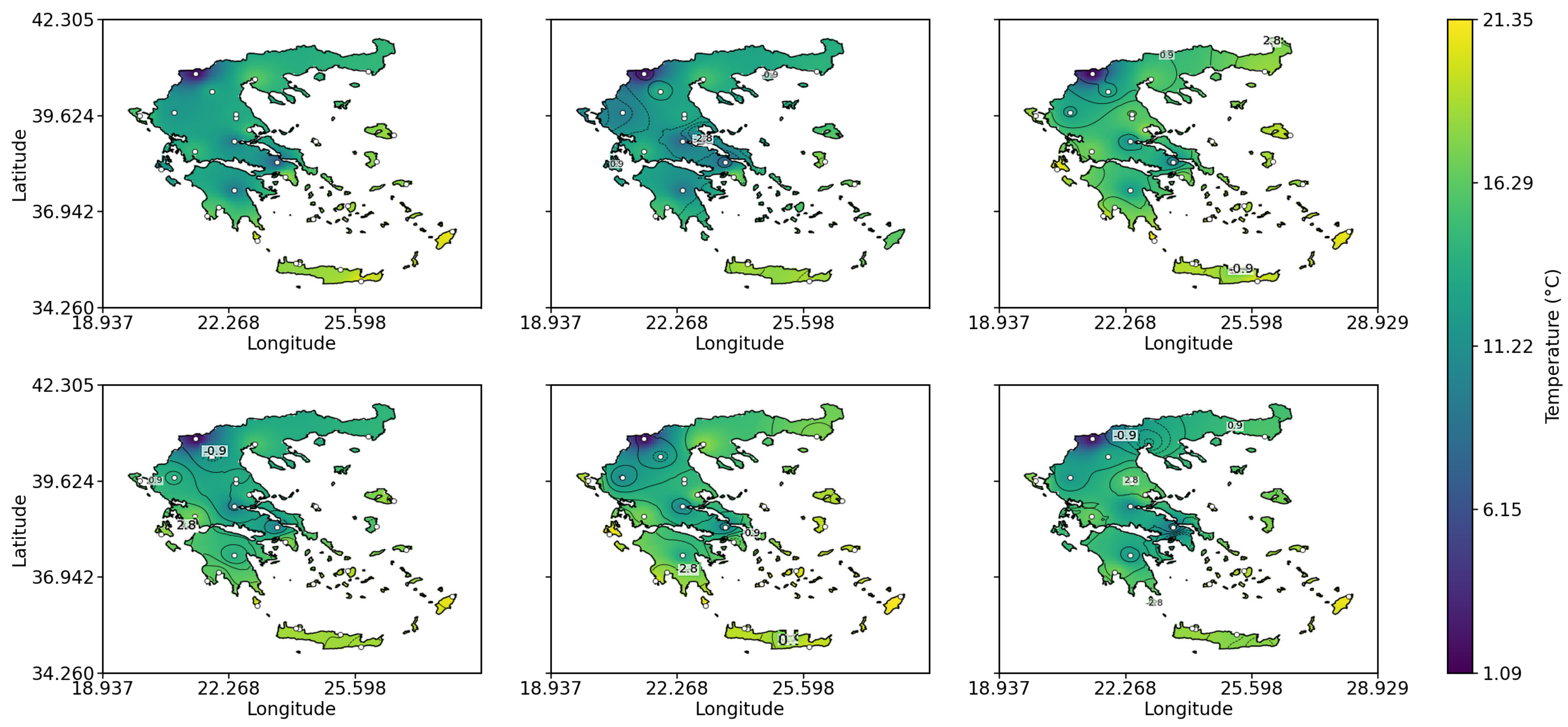

*Figure 6. Greece growing-season uncorrected TN spatial fields using IDW. The observed panel uses HNMS/EMY stations. The panels, read from left to right and top to bottom, show Observed, E-OBS, AgERA5, MARS-STAT/JRC Agri4Cast, ERA5, and ERA5-Land; contours show dataset-minus-observed differences; all temperatures are in °C.*

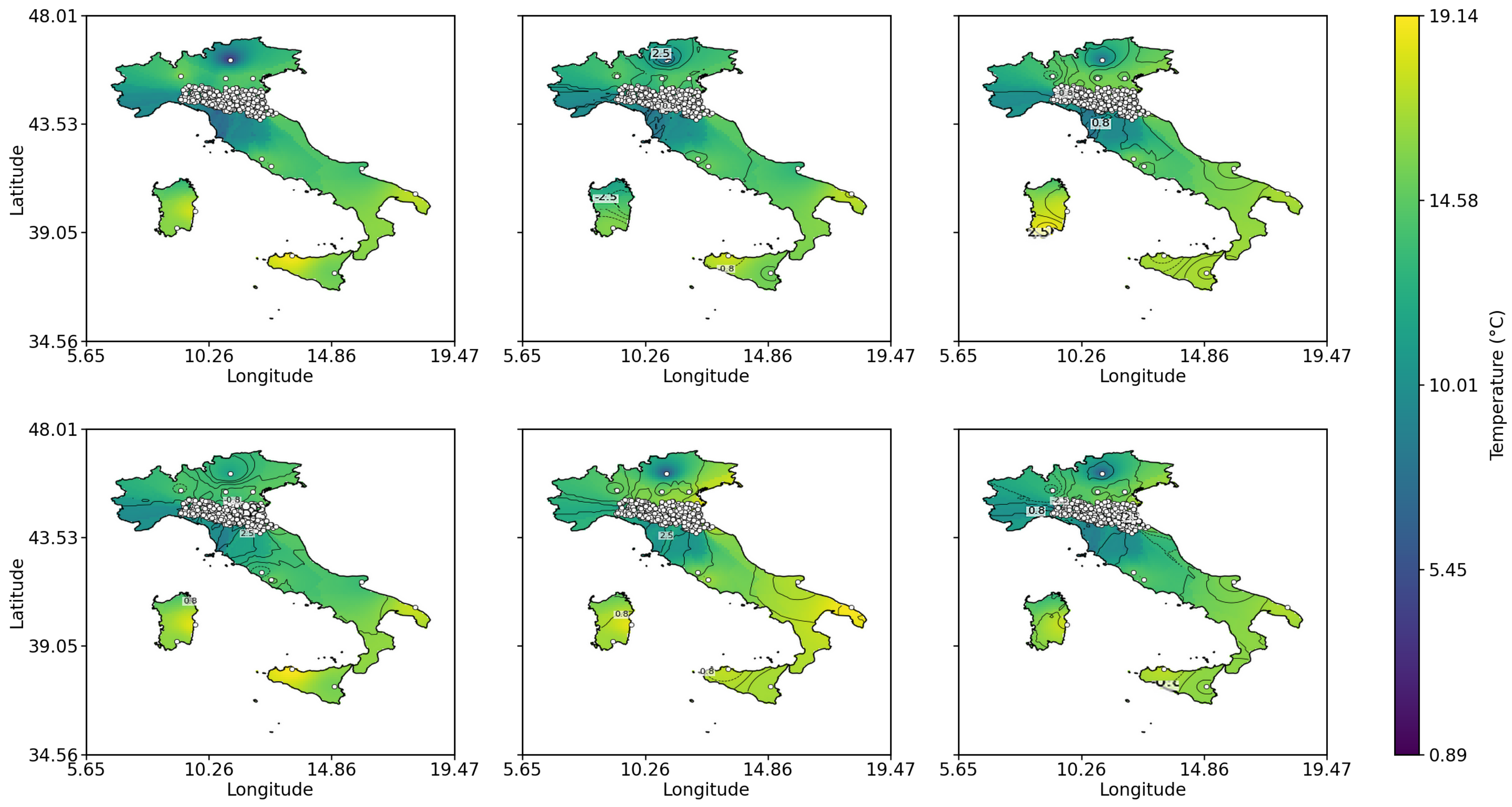

***Figure 7. Italy growing-season uncorrected TN spatial fields using IDW. The observed panel uses ECA&D stations. The panels, read from left to right and top to bottom, show Observed, E-OBS, AgERA5, MARS-STAT/JRC Agri4Cast, ERA5, and ERA5-Land; contours show dataset-minus-observed differences; all temperatures are in °C.***

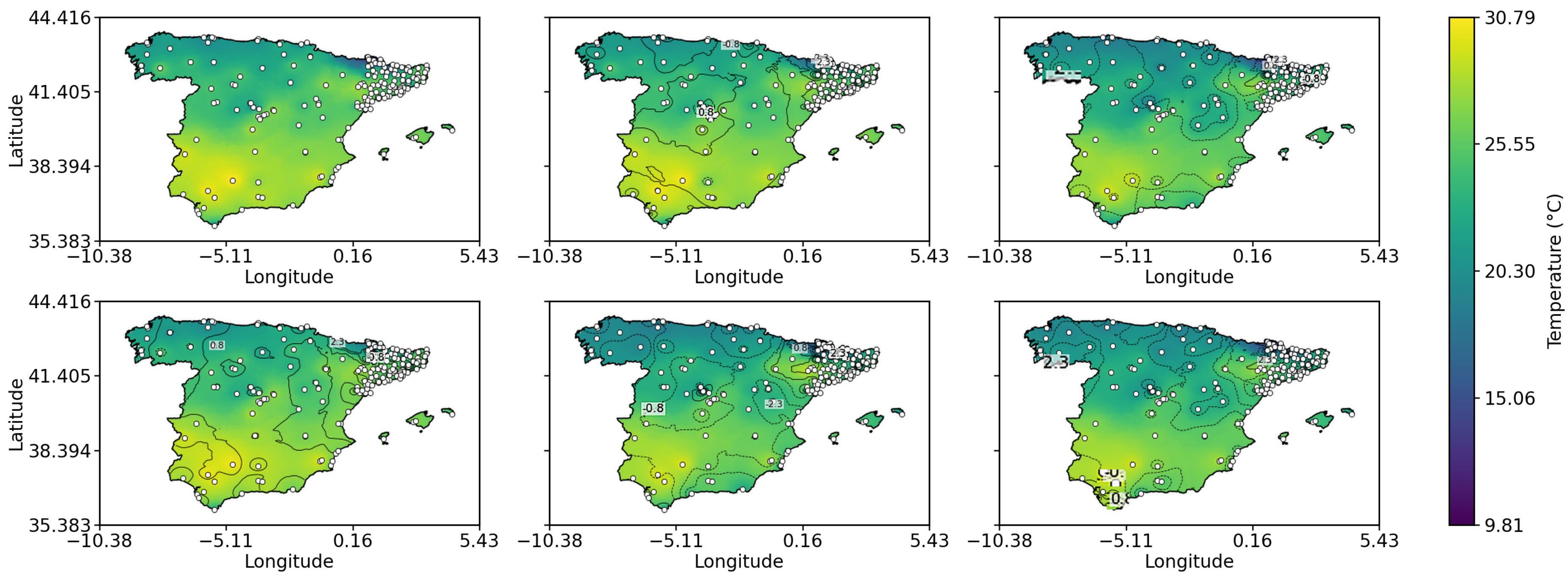

***Figure 8. Spain growing-season uncorrected TX spatial fields using IDW. The observed panel uses ECA&D stations. The panels, read from left to right and top to bottom, show Observed, E-OBS, AgERA5, MARS-STAT/JRC Agri4Cast, ERA5, and ERA5-Land; contours show dataset-minus-observed differences; all temperatures are in °C.***

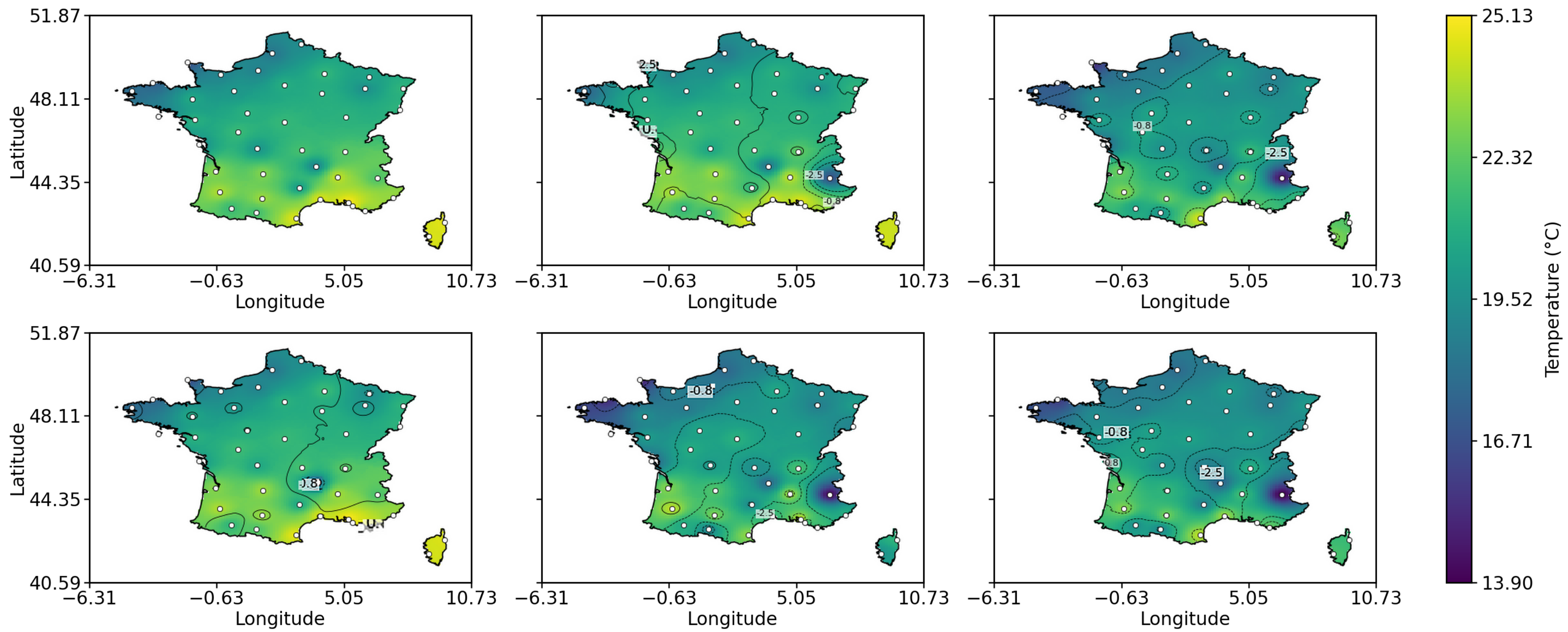

***Figure 9. France growing-season uncorrected TX spatial fields using IDW. The observed panel uses ECA&D stations. The panels, read from left to right and top to bottom, show Observed, E-OBS, AgERA5, MARS-STAT/JRC Agri4Cast, ERA5, and ERA5-Land; contours show dataset-minus-observed differences; all temperatures are in °C.***

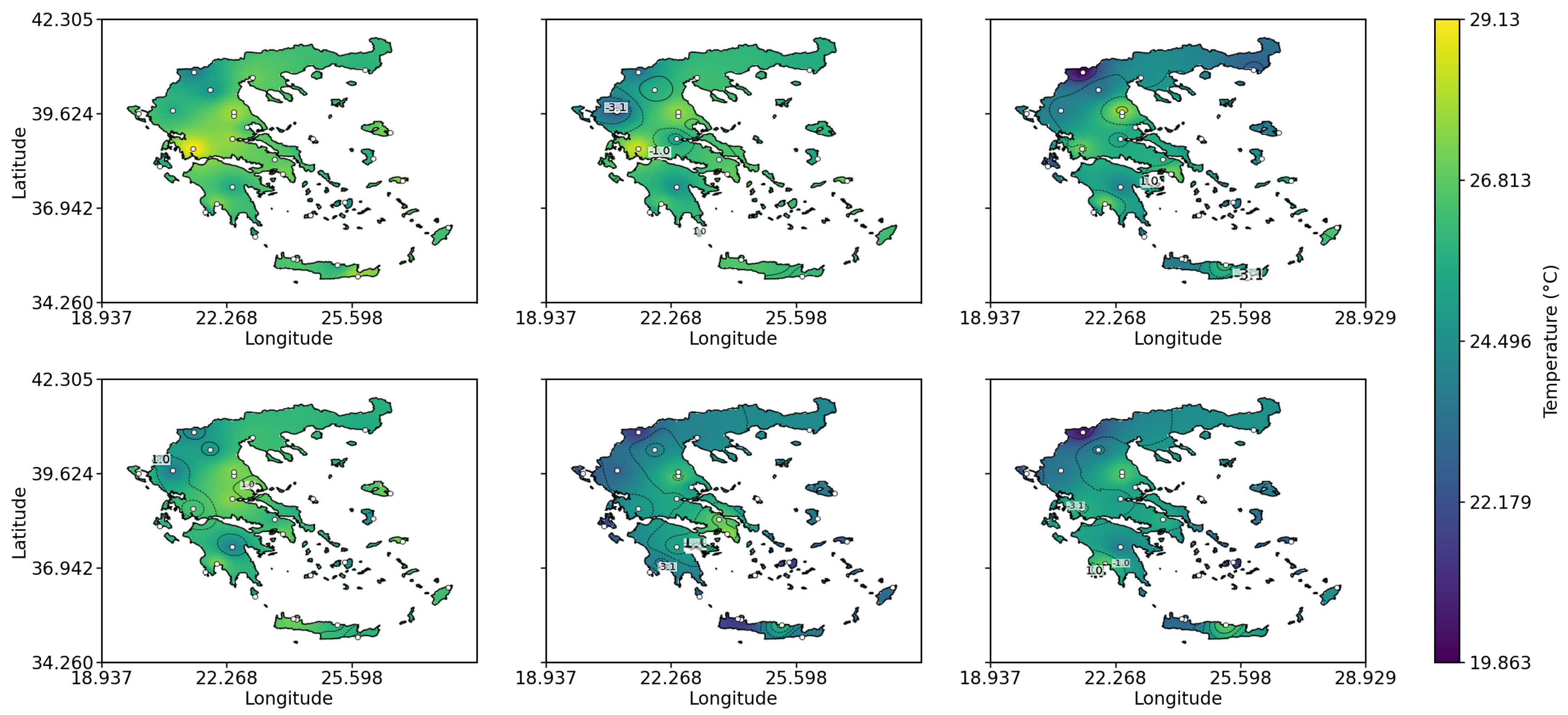

*Figure 10. Greece growing-season uncorrected TX spatial fields using IDW. The observed panel uses HNMS/EMY stations. The panels, read from left to right and top to bottom, show Observed, E-OBS, AgERA5, MARS-STAT/JRC Agri4Cast, ERA5, and ERA5-Land; contours show dataset-minus-observed differences; all temperatures are in °C.*

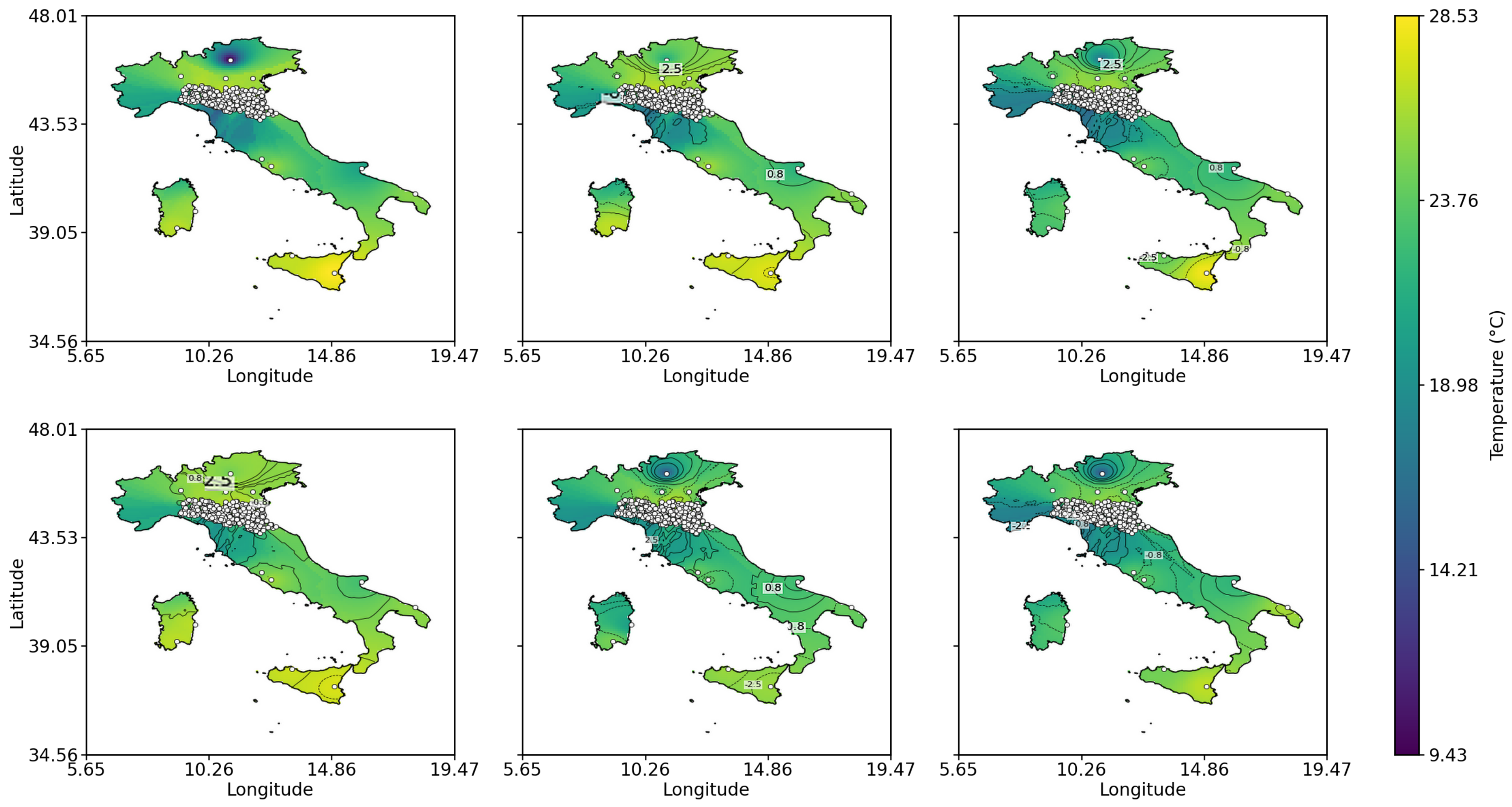

*Figure 11. Italy growing-season uncorrected TX spatial fields using IDW. The observed panel uses ECA&D stations. The panels, read from left to right and top to bottom, show Observed, E-OBS, AgERA5, MARS-STAT/JRC Agri4Cast, ERA5, and ERA5-Land; contours show dataset-minus-observed differences; all temperatures are in °C.*

### 3.5 Validation performance

Direct daily validation used identical support across all five datasets (Table 4; Figures 12, 13, and 14). E-OBS minimized the equally weighted mean of station-level RMSE values for TN, TX, Tmean, and DTR in all 16 country-variable comparisons. Its TN RMSE ranged from 0.418 °C in France to 1.211 °C in Italy, TX RMSE from 0.473 to 1.357 °C, Tmean RMSE from 0.371 to 0.978 °C, and DTR RMSE from 0.474 to 1.709 °C. Paired-bootstrap first-rank probabilities ranged from 0.79 to 1.00.

The all-five common-support analysis retained 159 stations in Spain, 37 in France, 13 in Greece, and 351 in Italy. Greek E-OBS TN and TX biases remained −0.407 and −0.663 °C, so the lowest RMSE did not imply negligible mean error. MARS-STAT/JRC Agri4Cast generally ranked second, whereas ERA5-derived datasets commonly underestimated TX and compressed DTR.

Monthly-climatology-removed anomalies confirmed that the annual cycle alone did not explain the daily ranking (Supplementary Figure S14). E-OBS again minimized TN, TX, Tmean, and DTR anomaly RMSE in all 16 comparisons. In the common-support Taylor diagrams, E-OBS lay nearest the observed reference, MARS-STAT/JRC Agri4Cast generally ranked next, and the ERA5-derived datasets more often showed reduced TX variability.

***Table 4. Best common-support daily dataset by country and variable. n denotes stations; bias and RMSE are equally weighted means of station-level values (°C). RMSE intervals and P(rank = 1) are from paired station bootstraps. Daily validation transformed station correlations with Fisher's z, averaged the transformed values with equal station weight, and back-transformed the mean to r.***

| Country | Metric | Best dataset | n | RMSE (95% CI) | Bias | Fisher-z-averaged r | P(rank = 1) |
|---|---|---|---|---|---|---|---|
| Spain | TN | E-OBS | 159 | 0.928 (0.804–1.058) | +0.102 | 0.9969 | 1.00 |
| Spain | TX | E-OBS | 159 | 1.154 (0.928–1.409) | +0.170 | 0.9984 | 1.00 |
| Spain | Tmean | E-OBS | 159 | 0.894 (0.733–1.072) | +0.136 | 0.9986 | 1.00 |
| Spain | DTR | E-OBS | 159 | 1.087 (0.931–1.255) | +0.067 | 0.9868 | 1.00 |
| France | TN | E-OBS | 37 | 0.418 (0.290–0.627) | −0.134 | 0.9990 | 1.00 |
| France | TX | E-OBS | 37 | 0.473 (0.287–0.765) | −0.125 | 0.9995 | 0.85 |
| France | Tmean | E-OBS | 37 | 0.371 (0.222–0.619) | −0.130 | 0.9996 | 0.93 |
| France | DTR | E-OBS | 37 | 0.474 (0.382–0.583) | +0.009 | 0.9964 | 1.00 |
| Greece | TN | E-OBS | 13 | 1.070 (0.666–1.515) | −0.407 | 0.9969 | 0.99 |
| Greece | TX | E-OBS | 13 | 1.142 (0.730–1.656) | −0.663 | 0.9988 | 0.94 |
| Greece | Tmean | E-OBS | 13 | 0.977 (0.590–1.444) | −0.535 | 0.9988 | 0.79 |
| Greece | DTR | E-OBS | 13 | 1.028 (0.707–1.334) | −0.256 | 0.9878 | 1.00 |
| Italy | TN | E-OBS | 351 | 1.211 (1.124–1.298) | +0.036 | 0.9931 | 1.00 |
| Italy | TX | E-OBS | 351 | 1.357 (1.238–1.490) | +0.036 | 0.9963 | 1.00 |
| Italy | Tmean | E-OBS | 351 | 0.978 (0.893–1.073) | +0.036 | 0.9969 | 1.00 |
| Italy | DTR | E-OBS | 351 | 1.709 (1.596–1.824) | +0.001 | 0.9641 | 1.00 |

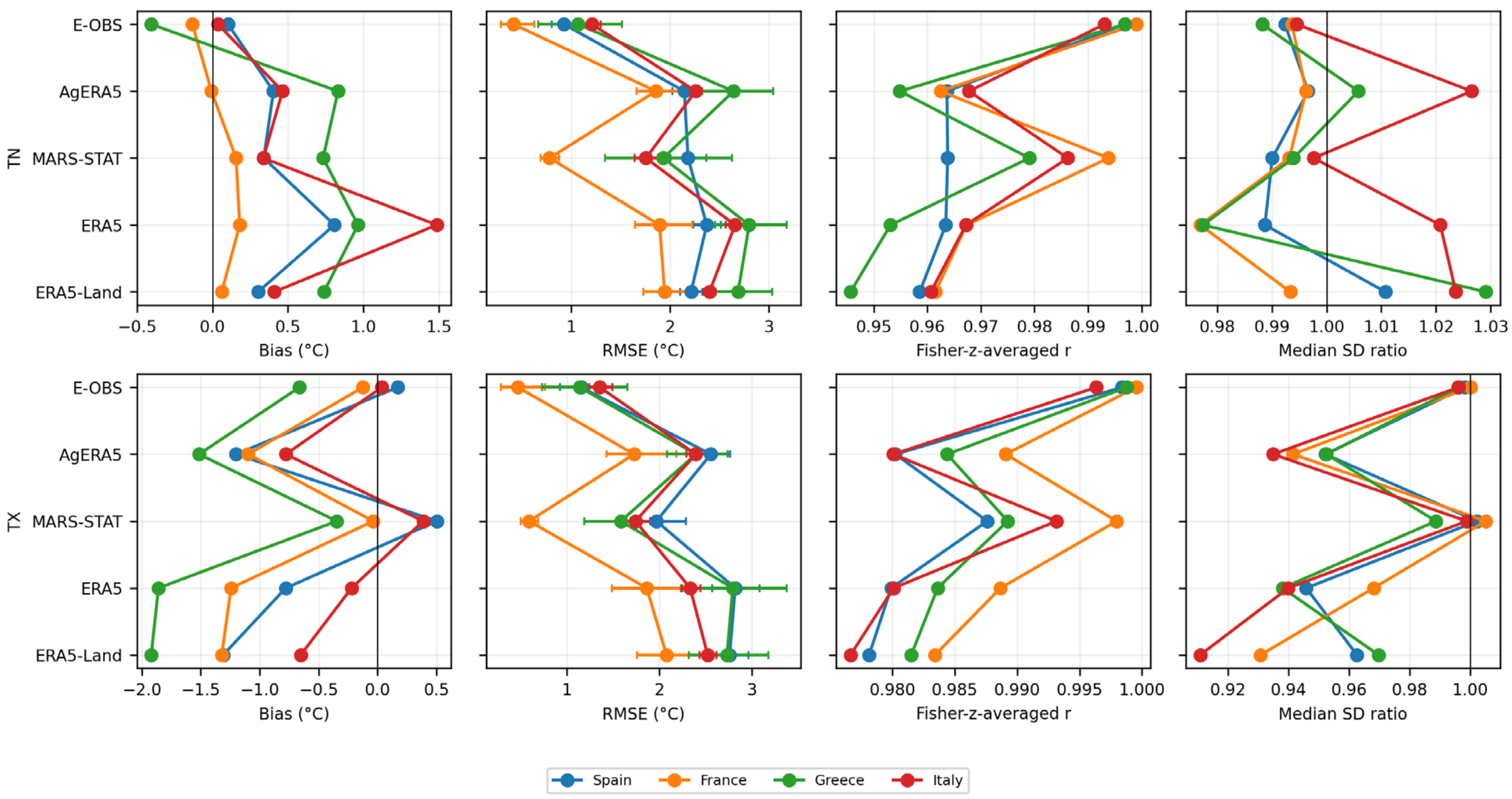

***Figure 12. Common-support daily validation by dataset, country, and variable. Columns show bias, root mean square error (RMSE), Fisher-z-averaged r, and median standard-deviation ratio. Daily validation transformed station correlations with Fisher's z, averaged the transformed values with equal station weight, and back-transformed the mean to r; error bars show paired station-bootstrap 95% RMSE intervals. TN, TX, Tmean, and DTR denote daily minimum temperature, daily maximum temperature, mean temperature, and diurnal temperature range; MARS-STAT denotes MARS-STAT/JRC Agri4Cast.***

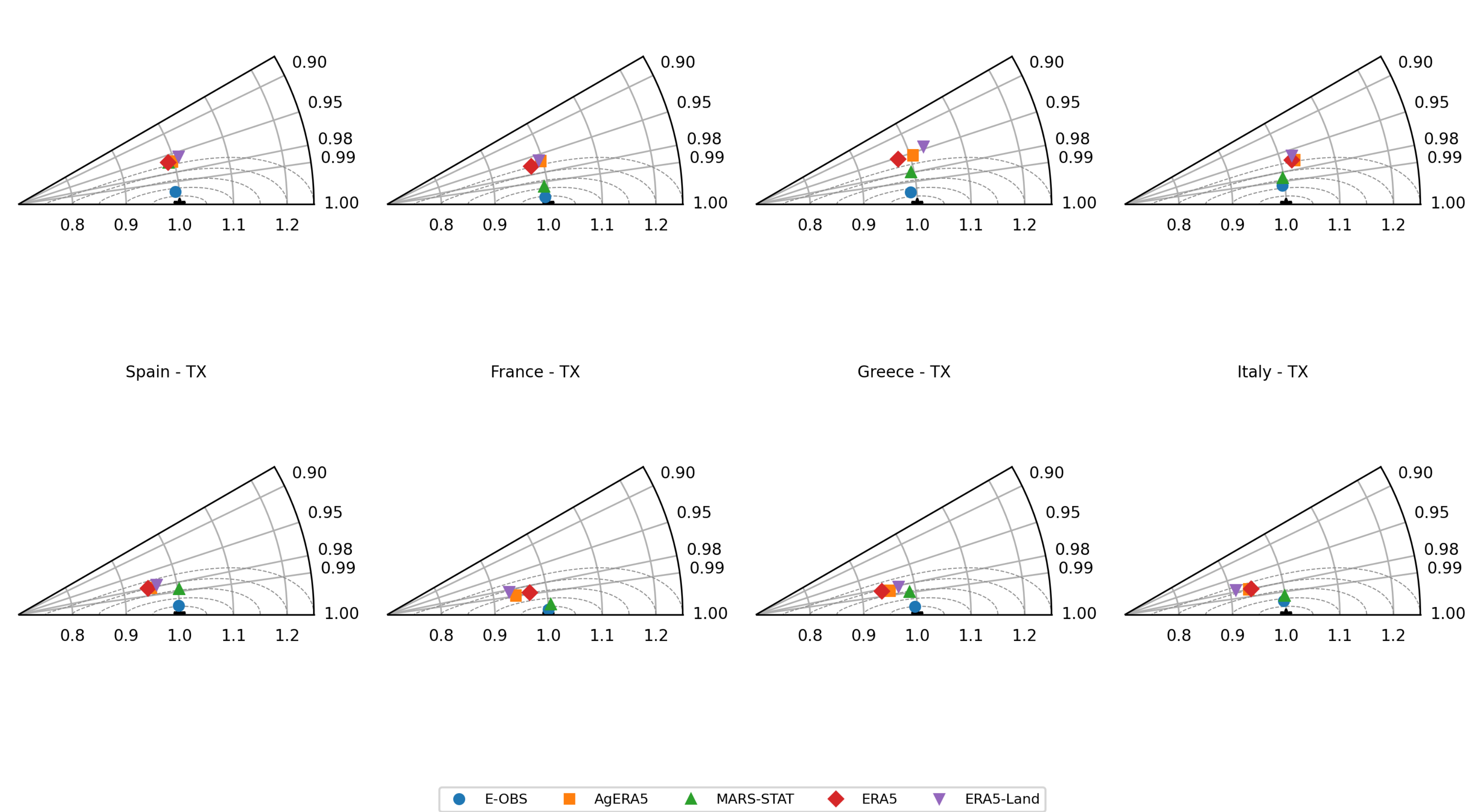

***Figure 13. Common-support Taylor synthesis for daily minimum (TN) and maximum (TX) temperature. The angular coordinate shows correlation, the radius shows the standard-deviation ratio, and dashed contours show normalized centered root mean square difference; the diagrams weight station metrics equally. MARS-STAT denotes MARS-STAT/JRC Agri4Cast.***

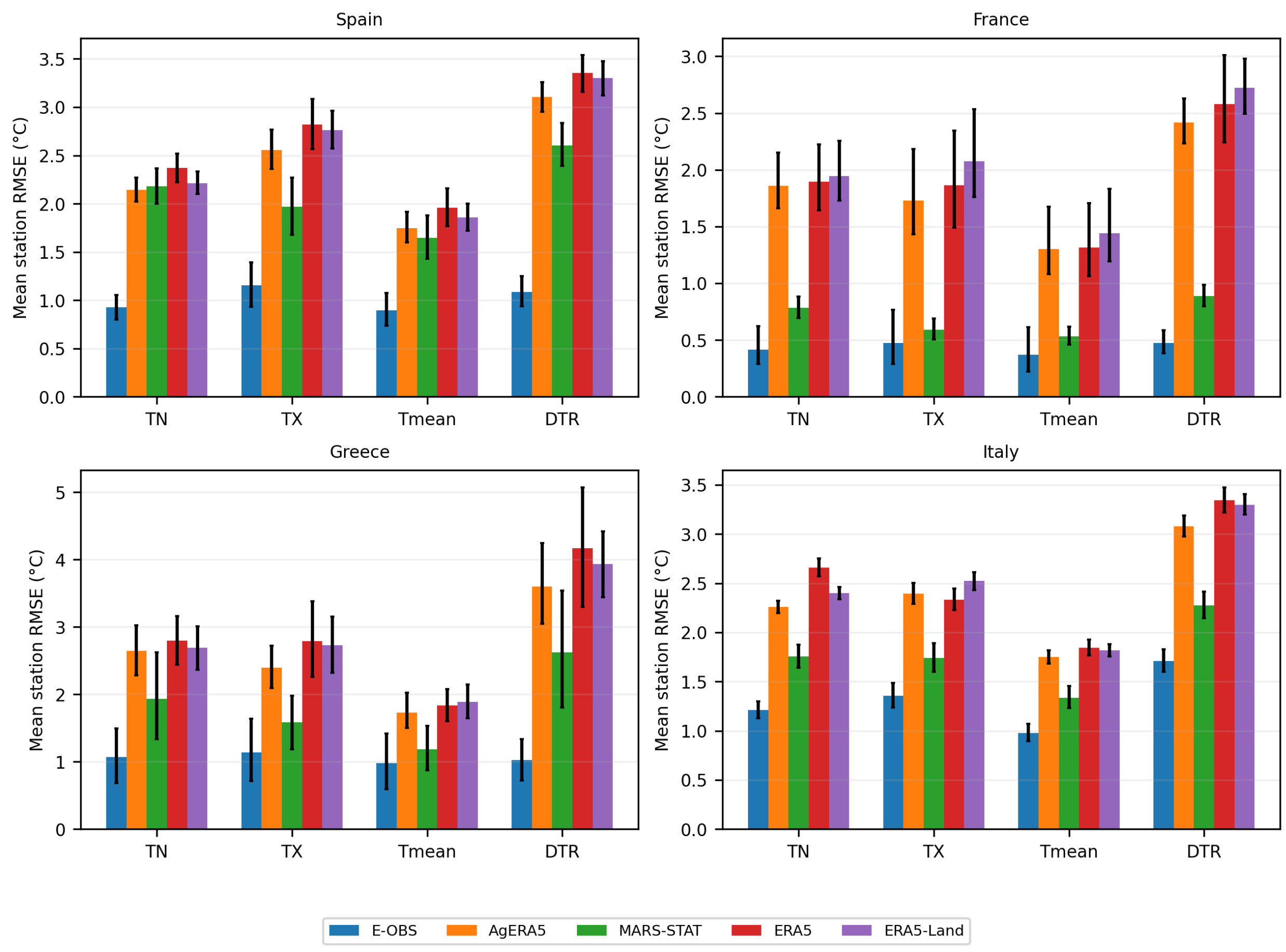

***Figure 14. Common-support daily root mean square error (RMSE) by country for daily minimum temperature (TN), daily maximum temperature (TX), mean temperature (Tmean), and diurnal temperature range (DTR). Error bars show paired station-bootstrap 95% intervals; MARS-STAT denotes MARS-STAT/JRC Agri4Cast.***

## 3.6 Growing-season GDD performance

Common-support GDD rankings differed from daily rankings (Table 5; Figure 15). E-OBS minimized mean station-level RMSE in Spain (129.0 degree-days) and Italy (123.9), whereas MARS-STAT/JRC Agri4Cast minimized it in France (46.1) and Greece (96.1). These RMSE values equaled 7.2%, 3.3%, 4.3%, and 7.1% of observed mean GDD in Spain, France, Greece, and Italy, respectively. Winkler-region agreement for the RMSE-leading datasets ranged from 71.2% in Italy to 91.6% in France. In France, however, E-OBS achieved slightly higher Winkler agreement than MARS-STAT/JRC Agri4Cast (92.8% versus 91.6%) despite MARS-STAT's lower RMSE.

Only the observed French GDD increase (+65.7 degree-days decade$^{-1}$) met the directional-robustness criterion: the HAC test remained significant after FDR control, the station-bootstrap interval excluded zero, and the fixed-network slope retained the positive sign. Spanish, Greek, and Italian country trends were network-sensitive. The GDD trend inference used all eligible observed station series (Spain n = 189, France n = 43, Greece n = 29, Italy n = 328), not the smaller common-support subsets used for performance ranking. The French GDD ranking remained uncertain: MARS-STAT/JRC Agri4Cast had a 0.55 paired-bootstrap probability of ranking first, whereas the first-rank probabilities were 0.82 in Greece and 1.00 in Spain and Italy.

***Table 5. Common-support growing-season GDD performance and observed trend robustness. The performance columns use identical all-five-dataset daily support and stations with at least five valid growing seasons; n (performance) denotes those stations. Winkler agreement is the mean station-level proportion of matched years assigned to the same classical region as the observed reference. The trend assessment uses all eligible observed GDD series (Spain n = 189, France n = 43, Greece n = 29, Italy n = 328). RMSE and bias are in degree-days. In France, E-OBS had slightly higher Winkler agreement (92.8%) than MARS-STAT/JRC Agri4Cast (91.6%) despite the latter's lower RMSE. Trend robustness follows the same directional criterion as Table 3.***

| Country | n (performance) | Best dataset | GDD RMSE (95% CI) | Bias | RMSE / observed mean | Winkler agreement for RMSE leader | P(rank = 1) | Trend assessment |
|---|---|---|---|---|---|---|---|---|
| Spain | 144 | E-OBS | 129.0 (104.4–156.4) | +9.6 | 7.2% | 77.0% | 1.00 | Sensitive |
| France | 36 | MARS-STAT/JRC Agri4Cast | 46.1 (32.4–61.0) | +7.8 | 3.3% | 91.6% | 0.55 | Directionally robust |
| Greece | 10 | MARS-STAT/JRC Agri4Cast | 96.1 (54.4–153.8) | +10.3 | 4.3% | 84.9% | 0.82 | Sensitive |
| Italy | 261 | E-OBS | 123.9 (109.0–141.0) | 0.0 | 7.1% | 71.2% | 1.00 | Sensitive |

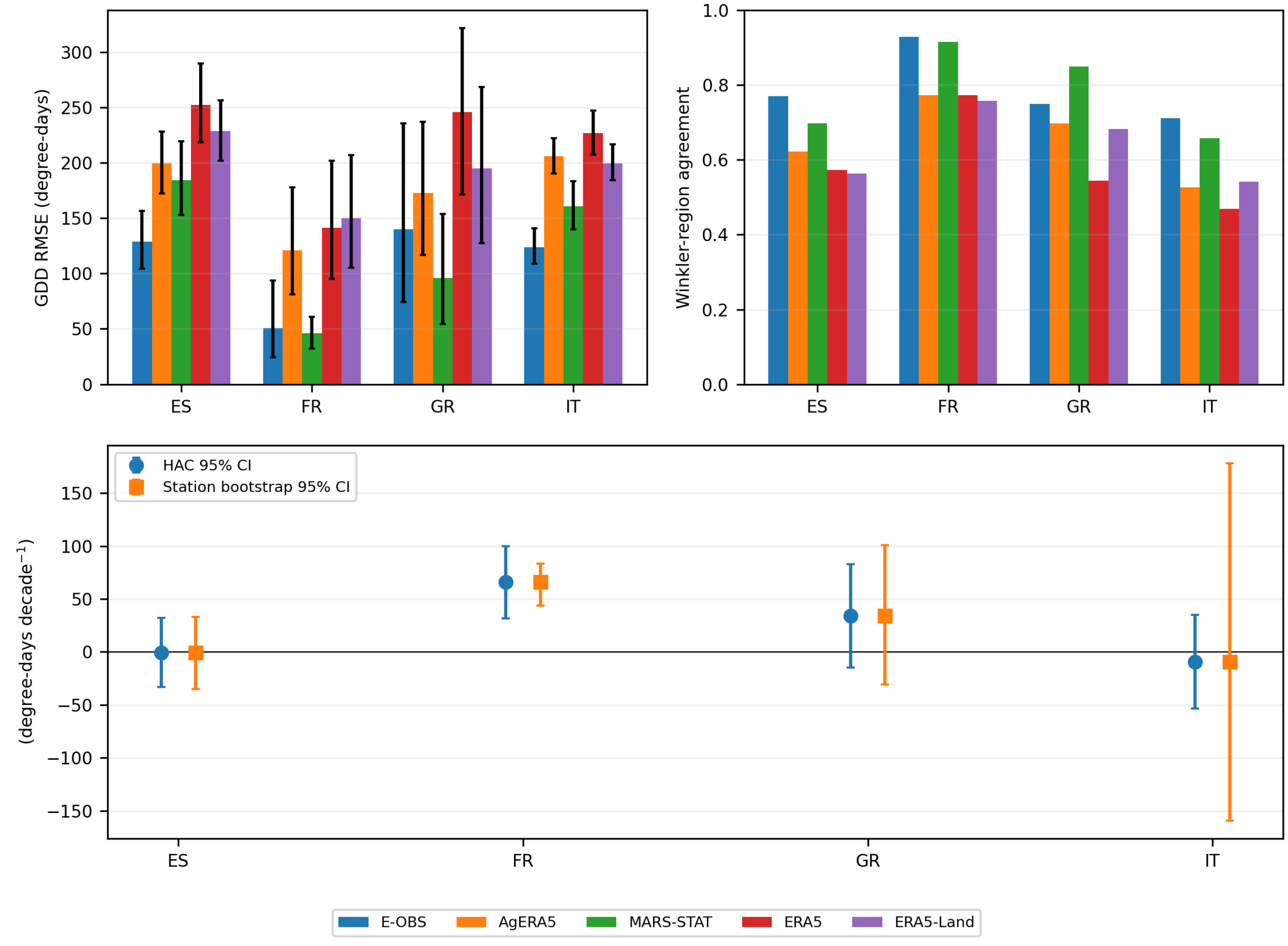

***Figure 15. Common-support growing degree days (GDD) RMSE and mean station-level Winkler-region agreement for stations with at least five valid growing seasons, together with observed GDD trend uncertainty. The two performance panels use common-support stations; the trend panel uses all eligible observed GDD series. The trend panel shows Newey–West heteroskedasticity- and autocorrelation-consistent (HAC) and station-bootstrap 95% intervals. ES, FR, GR, and IT denote Spain, France, Greece, and Italy; MARS-STAT denotes MARS-STAT/JRC Agri4Cast.***

## 3.7 Elevation-correction sensitivity

The fixed elevation correction had a systematic but variable effect (Table 6; Figure 16). It improved every TX comparison for E-OBS, AgERA5, ERA5, and ERA5-Land but generally degraded TN; MARS-STAT/JRC Agri4Cast showed a mixed TX response and no TN improvement. Overall, 22 of 40 comparisons improved, so the correction remains diagnostic rather than a default post-processing step.

*Table 6. Summary of the elevation-correction sensitivity by dataset and temperature variable. ΔRMSE is corrected RMSE minus uncorrected RMSE; mean and median ΔRMSE summarize the four country-specific changes; negative values indicate improvement. MARS-STAT denotes MARS-STAT/JRC Agri4Cast.*

| Dataset | Variable | Country comparisons | Improved | Mean ΔRMSE (°C) | Median ΔRMSE (°C) |
|---|---|---|---|---|---|
| AgERA5 | TN | 4 | 0 | 0.267 | 0.281 |
| AgERA5 | TX | 4 | 4 | −0.312 | −0.264 |
| E-OBS | TN | 4 | 1 | 0.153 | 0.191 |
| E-OBS | TX | 4 | 4 | −0.210 | −0.212 |
| ERA5 | TN | 4 | 2 | 0.108 | 0.055 |
| ERA5 | TX | 4 | 4 | −0.578 | −0.487 |
| ERA5-Land | TN | 4 | 1 | 0.151 | 0.164 |
| ERA5-Land | TX | 4 | 4 | −0.455 | −0.444 |
| MARS-STAT | TN | 4 | 0 | 0.310 | 0.294 |
| MARS-STAT | TX | 4 | 2 | 0.060 | 0.060 |

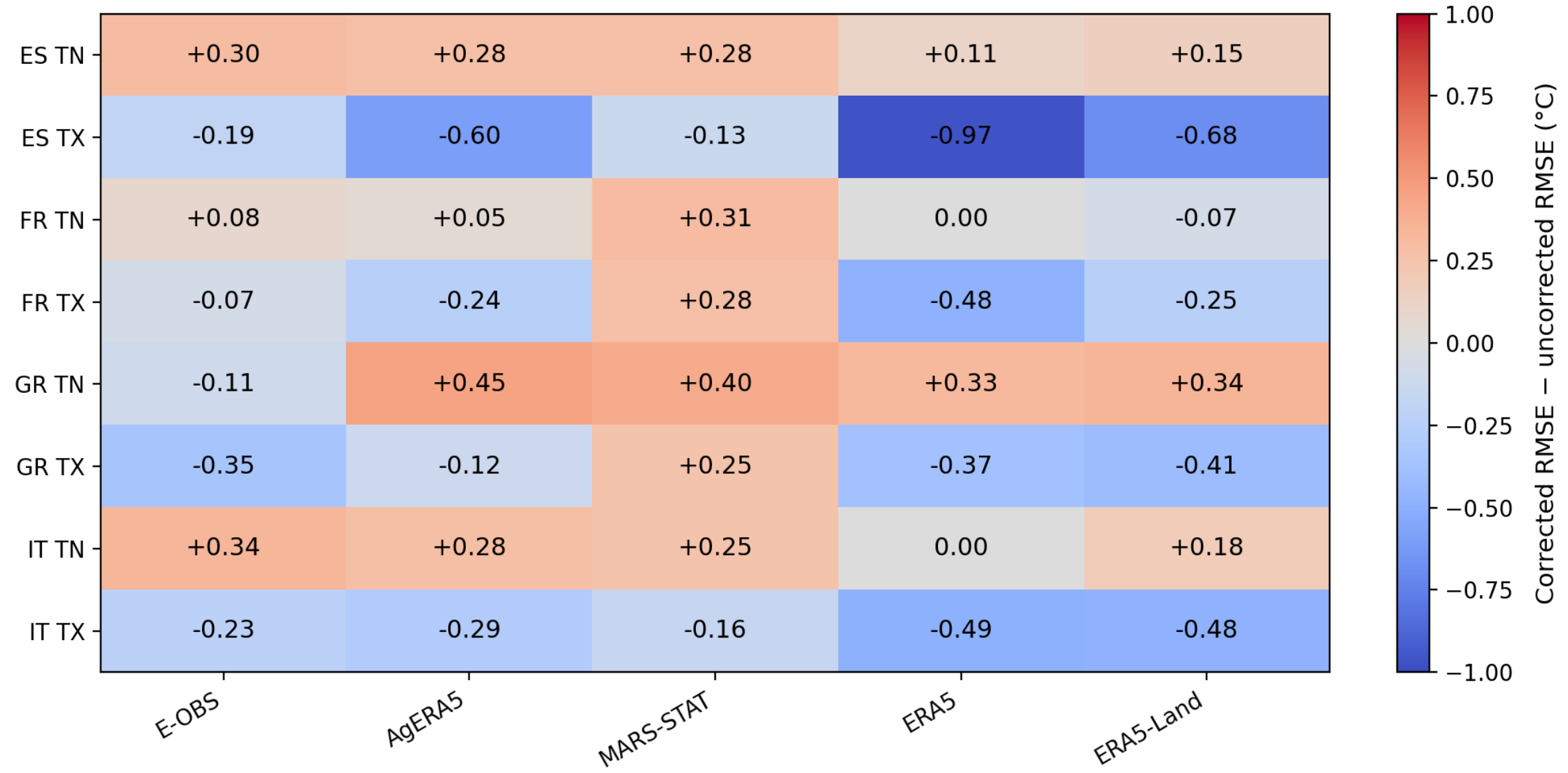

*Figure 16. Change in station-level RMSE after the fixed elevation correction, calculated as corrected RMSE minus uncorrected RMSE (°C). Negative values indicate improvement; the uncorrected comparison remains primary. ES, FR, GR, and IT denote Spain, France, Greece, and Italy; MARS-STAT denotes MARS-STAT/JRC Agri4Cast.*

# 4. Discussion

## 4.1 Performance across datasets and timescales

Dataset rankings changed with support and decision metric. On identical daily support, E-OBS led all 16 TN, TX, Tmean, and DTR comparisons and all 16 anomaly comparisons. Under the 10-year seasonal threshold, E-OBS led 42 of 64 RMSE comparisons and MARS-STAT/JRC Agri4Cast led 22. GDD favored E-OBS in Spain and Italy and MARS-STAT/JRC Agri4Cast in France and Greece, while most trend targets were themselves network-sensitive.

E-OBS's strong daily performance is consistent with its station-based construction but is not observation-independent. Overlap between construction and evaluation networks can raise apparent agreement, while changing station support affects smoothing, distributions, and trends [2], [8], [9]; tail errors also affect climate-model evaluation [10]. Low RMSE therefore indicates close agreement with the available stations, not absence of structural uncertainty.

Construction mechanism alone did not determine performance. E-OBS agreed well in Spain, consistent with Iberia01 [11], but still showed climatological offsets; in mountainous, island-rich Greece, it produced the lowest daily RMSE while its monthly TN and TX climatologies remained cool-biased [12]. Rankings therefore remained statistic- and network-dependent.

MARS-STAT/JRC Agri4Cast was usually second for daily variables, closely reproduced the French monthly cycle, and minimized GDD error in France and Greece. It was not uniformly superior in Spain or Italy, nor did it consistently reproduce trends. The earlier Greek precipitation study likewise found season- and index-dependent rankings [20], so researchers should not transfer performance rankings between variables.

### 4.2 Reanalysis coherence, amplitude, and DTR

The ERA5-derived datasets captured the broad temporal sequence more faithfully than they reproduced daily amplitude. Their high correlations coexisted with cool TX bias and compressed DTR. The anomaly analysis reached the same ranking after removing station-dataset monthly climatologies, showing that the result was not only an artifact of the annual temperature cycle.

A Europe-wide ERA5 evaluation likewise found strong annual-cycle agreement but larger errors over the Alps and Mediterranean [13]. The common-support anomaly results support the same interpretation: correlation, bias, variance, and weather-scale RMSE answer different questions. Figure 13 therefore complements Figures 12 and 14 rather than serving as an independent ranking.

ERA5-Land did not consistently improve on ERA5. Finer grids represent land-surface heterogeneity more explicitly but cannot remove forcing, radiation, boundary-layer, or exposure errors [5], [14]. AgERA5 showed similar coherence but still required local validation in Greece [24]. Resolution alone did not guarantee station-level accuracy.

DTR sharpened these contrasts because it combines TN and TX errors. Interpolation and reanalysis can preserve warm/cool sequencing while damping diurnal amplitude [9], [10]. Thermal-range applications should therefore validate DTR directly.

### 4.3 Spatial representation, terrain, and station density

Figures 4–11 show that datasets reproduced broad country gradients more consistently than local structure in mountains, coastal zones, and islands. Greece's 29-station HNMS/EMY network sampled elevations only up to 662 m, so the study draws no conclusions about the country's highest terrain. Spatial-weighting and annual network-composition analyses further show that low national RMSE can coexist with local and sampling-dependent departures.

Departures concentrated around relief and land-sea boundaries, consistent with national evaluations. Iberia01 showed spatially heterogeneous differences from E-OBS, while CLIMADAT-GRid showed that higher local resolution and station support improved Greek mountain and island representation [11], [12]. Across the four countries, broad gradients were robust, but the magnitude and sign of local departures depended on terrain, coastlines, station density, and dataset construction.

IDW provided a consistent visualization of station evidence rather than an assumed optimal interpolation. Skill depends on network density, elevation, season, and high-elevation coverage, and no simple method dominates all settings [23]. The maps locate broad biases for comparison with station metrics; they are descriptive summaries, not new gridded climate fields or evidence independent of the interpolated observations.

### 4.4 Country-level and local station trends

Country-average trend inference was substantially less stable than unadjusted OLS p values suggested. Among annual changes, only Spanish DTR and French TX satisfied the HAC test after FDR control, the station-bootstrap interval criterion, and the fixed-network sign criterion. French TN and Tmean warming remained directionally consistent but missed at least one threshold. Estimates for Spanish TN and Tmean and for Italian TN and DTR changed materially under fixed-network or spatial-weighting sensitivity. Researchers should therefore condition trend-dataset rankings on the robustness of the observed target.

Researchers have used E-OBS for continental temperature monitoring [25], but these country series are not a homogenized regional index. Changing networks can alter variance and trends [9], and periods, station availability, and TN and TX record lengths differed among countries. The station-balanced series therefore represent country-specific evidence, not a common Mediterranean trend.

In Greece, the weak positive TN slope and negative TX slope are qualitatively consistent with the stronger trend in TN than in TX reported at viticultural sites [18]. Because stations and periods differ and the present slopes are nonsignificant, the comparison is qualitative; it nevertheless shows why analysts should not collapse TN and TX into Tmean.

Station-specific trends provide a complementary local scale, not a substitute for network-robust country series (Supplementary Figures S1–S3 and S15–S17). Positive local medians in Spain and Italy coexist with network-sensitive country estimates because stations have different windows and spatial weights. The Methods describe the potential-breakpoint screen; Supplementary Figure S13 reports only annual trend-uncertainty diagnostics.

Local trend accuracy differed from daily and seasonal-mean accuracy (Supplementary Figures S4 and S5). E-OBS minimized Tmean and GDD slope MAE in Spain, France, and Italy, whereas MARS-STAT/JRC Agri4Cast had the lowest MAE in Greece. Supplementary Figures S4 and S5 show all retained station points, and researchers should restrict comparisons to the same record windows.

## 4.5 GDD, seasonal timing, and agricultural implications

Table 5 combines common-support GDD error, paired-bootstrap uncertainty, observed-mean scaling, Winkler-region agreement, and directional trend robustness. Table 7 translates the full evidence into a fitness-for-use matrix. The frost- and hot-day sensitivities (Supplementary Figure S18) generally favored E-OBS, with MARS-STAT/JRC Agri4Cast leading both French threshold diagnostics.

***Table 7. Application-specific fitness-for-use matrix based on the main and supplementary robustness analyses. MARS-STAT denotes MARS-STAT/JRC Agri4Cast.***

| Application | Primary evidence | Leading dataset(s) | Main caveat |
|---|---|---|---|
| Daily TN/TX/Tmean/DTR | Identical station-day support across all five datasets; paired station bootstrap | E-OBS in all 16 country-variable comparisons | Potential station-network dependence; source calendar-day labels retained |
| Daily anomalies | Monthly-climatology-removed common support | E-OBS in all 16 comparisons | Weather-scale fidelity remains subject to point-grid mismatch |
| Seasonal means | At least 10 matched years; five-year sensitivity reported | E-OBS in Spain and Italy; MARS-STAT led most comparisons in France; mixed in Greece | Greek primary subset contains four stations; rankings are threshold-sensitive |
| Spatial fields | Uncorrected TN/TX fields and station-support diagnostics | E-OBS generally follows broad gradients | IDW is descriptive; high mountains, coasts, and islands remain undersampled |
| Growing-season GDD | Identical daily support, paired bootstrap, Winkler-region agreement | E-OBS in Spain and Italy; MARS-STAT in France and Greece | Near-tie in France; thermal classes do not represent phenology |
| Trend analysis | HAC, station bootstrap, fixed network, FDR, and spatial weighting | No universal gridded-dataset leader | Most observed slopes are network- or inference-sensitive |
| Elevation adjustment | Uncorrected-versus-fixed-lapse-rate sensitivity | Diagnostic only | Usually helps TX but often worsens TN |
| Frost- and hot-day sensitivity | Common-support April–May TN < 0 °C and June–August TX ≥ 35 °C | E-OBS generally; MARS-STAT in France | Operational thresholds, not cultivar-specific phenological endpoints |

GDD changed the ranking because it accumulates only the positive daily excess of Tmean over the 10 °C base, and threshold conventions alter the resulting index [17]. Persistent near-threshold errors can therefore change annual rankings even when mean temperature bias is small.

The seasonal analysis identifies within-season timing that April–October GDD alone cannot reveal. France showed directionally robust warming in spring and across the growing season, with the strongest point estimates in spring. Greece retained directionally robust summer Tmean warming. Italian growing-season, spring, and summer DTR slopes met the directional criterion, although the fixed-network slopes were small and nonsignificant. The April–May frost and June–August hot-day counts are operational sensitivity indicators, not observations of grapevine phenology.

#### 4.5.1 Spain

In Spain, E-OBS minimized common-support GDD RMSE (129.0 degree-days; 7.2% of observed mean GDD) and achieved 77.0% Winkler-region agreement. The country GDD trend was network-sensitive despite positive local slopes. Together with the north-south and elevation contrasts in Figures 4 and 8, these results support E-OBS as the best-supported national-scale dataset evaluated, but they do not justify unqualified vineyard-scale zoning.

#### 4.5.2 France

France had the smallest GDD error. MARS-STAT/JRC Agri4Cast produced the lowest RMSE (46.1 degree-days) and the lowest frost- and hot-day RMSEs, whereas E-OBS had slightly higher Winkler-region agreement (92.8% versus 91.6%). MARS-STAT's probability of ranking first for GDD was 0.55 because E-OBS performed nearly as well. Directionally robust spring and growing-season warming, together with the north–south and Corsican contrasts in Figures 5 and 9, underscores the importance of early-season heat accumulation.

#### 4.5.3 Greece

Greece combined a pronounced mainland-island contrast with the smallest common-support samples. MARS-STAT/JRC Agri4Cast minimized GDD RMSE (96.1 degree-days; 4.3% of observed mean) and achieved 84.9% Winkler-region agreement, but the primary seasonal analysis contained only four stations; the daily and GDD analyses retained 13 and 10 stations, respectively. Figures 6 and 10 therefore support regional comparisons only within the sampled elevation range of 0–662 m; they do not support inference for the highest mountains or individual vineyards.

#### 4.5.4 Italy

In Italy, E-OBS minimized common-support GDD RMSE (123.9 degree-days; 7.1% of observed mean) and achieved 71.2% Winkler-region agreement. The country GDD and annual temperature trends were network-sensitive, while many local records showed warming. The Alpine, Apennine, southern mainland, Sicilian, and Sardinian contrasts in Figures 7 and 11 therefore argue for site-specific checks even when the national error ranking is stable.

#### 4.5.5 Cross-country viticultural interpretation

Across countries, the lowest GDD RMSE represented 3.3–7.2% of observed mean GDD. Winkler-region agreement for the RMSE-leading datasets ranged from 71.2% to 91.6%, quantifying the potential for thermal-zone misclassification. The reported expansion of warmer viticultural zones [19] is consistent with the robust French warming signal, but unequal periods and network sensitivity preclude a single Mediterranean trend; Greek evidence also emphasizes mainland, coastal, island, and elevation contrasts [18].

### 4.6 Elevation correction and representativeness

The fixed lapse-rate adjustment improved 22 of 40 comparisons, but its effect depended strongly on the temperature variable. It improved TX for E-OBS and the three ERA5-derived datasets and generally worsened TN; MARS-STAT/JRC Agri4Cast showed no TN improvement and a mixed TX response. This asymmetry indicates that elevation mismatch contributed to TX error but was not the dominant control on local TN.

Elevation corrections can markedly reduce reanalysis error at some high-altitude stations, yet fixed lapse rates are unreliable during particular seasons and at locations where local atmospheric controls dominate [15]. An ERA5-based study likewise found no single lapse-rate formulation that was consistently superior for daily minima and maxima, especially under inversions [16]. The present TN/TX contrast supports retaining the uncorrected comparison as primary and using the corrected values only to diagnose terrain-related sensitivity.

Elevation is only one component of point-grid representativeness. Land cover, exposure, urban influence, coastline placement, boundary-layer regime, and unresolved topography can differ even at equal elevations. A fixed lapse-rate correction is therefore not a universal post-processing recipe.

### 4.7 Implications for dataset selection

Dataset selection should begin with the decision variable and support rule (Table 7). Daily and anomaly applications require bias, RMSE, correlation, variance, and DTR diagnostics on paired support; seasonal applications require a stated minimum record length; spatial use requires network and terrain checks; trend studies require directionally robust observed targets; and viticulture requires direct GDD, Winkler-region agreement, and threshold diagnostics.

No dataset was uniformly best. E-OBS gave the closest common-support daily and anomaly agreement, MARS-STAT/JRC Agri4Cast led selected French and Greek seasonal and GDD comparisons, and trend fidelity had no stable dataset hierarchy. As in the precipitation study [20], choice depends on variable, support, season, terrain, and application. The archive does not retain the exact overlap between the construction and evaluation networks for E-OBS or MARS-STAT/JRC Agri4Cast, nor the hourly metadata needed to test local-day boundaries; the Supplementary Methods describe these unresolved dependencies.

## 5. Conclusions

This four-country evaluation separated common-support daily agreement, anomaly fidelity, seasonal performance, spatial representation, trend robustness, and agroclimatic accumulation. E-OBS led all daily and anomaly RMSE comparisons and most seasonal comparisons; MARS-STAT/JRC Agri4Cast led selected seasonal and GDD metrics in France and Greece. Only a small subset of annual and seasonal observed trends remained directionally robust after temporal-dependence adjustment, station resampling, and fixed-network sensitivity testing.

Dataset selection should therefore be conditional rather than universal. Direct rankings require identical support, trend claims require a stable observed target, and agricultural applications require GDD and threshold diagnostics rather than temperature RMSE alone. The fixed lapse-rate test remains diagnostic because it often improved TX but worsened TN.

## Supplementary Materials

Figures S1–S18 accompany this article. They provide local TN, TX, Tmean, DTR, and GDD trends, seasonal robustness, network-composition diagnostics, common-support daily-anomaly performance, viticulture-oriented threshold sensitivity, and trend-robustness evidence. The Supplementary Methods document the common-support rules, uncertainty procedures, quality-control checks, known provenance limits, and time-boundary limitations.

## Declaration of Generative AI and AI-Assisted Technologies in the Writing Process

During the preparation of this work, the author used OpenAI’s ChatGPT to explore improvements to the text and to assist with drafting and reviewing deterministic analysis and document-processing scripts. The author independently reviewed the scripts, verified the resulting statistics and figures against the underlying evidence, and edited the manuscript as needed. The author takes full responsibility for the content of the publication.

## Funding

This research received no external funding.

## Institutional Review Board Statement

Not applicable.

## Informed Consent Statement

Not applicable.

## Data Availability Statement

The observed reference uses ECA&D records for Spain, France, and Italy and HNMS/EMY records for Greece. Provider access conditions govern HNMS/EMY observations, and the author does not redistribute them. The original providers make the gridded datasets available under their respective access terms. The article and Supplementary Materials report the principal numerical summaries and robustness diagnostics. No public data or code repository accompanies this article; the author retains the underlying analytical files.

## Acknowledgments

The author acknowledges the Hellenic National Meteorological Service (HNMS; EMY) for the Greek station observations; the data providers and maintainers of ECA&D and E-OBS; the Copernicus Climate Data Store and ECMWF; and the European Commission Joint Research Centre. The author also acknowledges the Alexander S. Onassis Public Benefit Foundation for a personal doctoral scholarship supporting living expenses. This scholarship did not fund the research reported in this article.

## Conflicts of Interest

The author declares no conflict of interest.

# Supplementary Materials

## Evaluating E-OBS, AgERA5, MARS-STAT, ERA5, and ERA5-Land for Daily Minimum and Maximum Temperature Across Four Mediterranean European Countries

**Dimitrios Voulanas**

Texas A&M Energy Institute, Texas A&M University, College Station, TX, USA

Correspondence: dvoulanas@tamu.edu

**Abbreviations: TN, daily minimum temperature; TX, daily maximum temperature; Tmean, daily mean temperature; DTR, diurnal temperature range; GDD, growing degree days; ECA&D, European Climate Assessment & Dataset; HNMS/EMY, Hellenic National Meteorological Service; MARS-STAT, MARS-STAT/JRC Agri4Cast.**

## Supplementary Methods

Observed-data providers and support. Spain, France, and Italy used ECA&D station observations; Greece used Hellenic National Meteorological Service (HNMS; EMY) observations. Greek TX source availability extended through 2012, whereas paired TN–TX analyses ended in 2004 because the corresponding TN record ended there. Daily, daily-anomaly, seasonal-mean, GDD, Winkler-region, frost-day, and hot-day rankings used identical stations and dates across the observed reference and all five gridded datasets. Monthly climatology, trend, and spatial analyses used analysis-specific support reported in the relevant captions.

Daily and anomaly validation. The study calculated common-support daily and anomaly metrics at each station and then summarized them with equal station weight by country. Paired station bootstraps provided 95% intervals and first-rank probabilities. Removing monthly climatologies separated weather-scale departures from agreement driven by the annual temperature cycle.

Seasonal validation. The primary growing-season, spring, summer, and autumn analysis required at least 10 matched years per station; a five-year threshold provided sensitivity evidence. The seasonal analysis transformed station correlations with Fisher's z, averaged the transformed values with equal station weight, and back-transformed the mean to r. The median standard-deviation ratio and median absolute natural log-ratio summarized variability; a ratio of 1 and an absolute log-ratio of 0 indicate exact agreement. Only four Greek stations met the primary 10-year criterion, so the interpretation also considered the five-year sensitivity results.

Trend robustness and network sensitivity. Trend analysis evaluated country-year slopes using ordinary least squares (OLS), Newey–West heteroskedasticity- and autocorrelation-consistent (HAC) intervals, station-bootstrap intervals, Benjamini–Hochberg false discovery rate (FDR) control, and a fixed network. The fixed network required valid values in at least 80% of years and at least one valid value in both the first and last 20% of the country record. The maximum HAC lag was max(1, floor[4(T/100)^(2/9)]), where T denotes the number of annual values. One-degree spatial-cell weighting, elevation-tertile estimates, the Spanish mainland-and-Balearic map domain, and Pettitt potential-breakpoint screening provided additional sensitivity evidence. The study classified a trend as directionally robust only when the HAC test remained significant after FDR control, the station-bootstrap interval excluded zero, and the fixed-network slope retained the same sign. Fixed-network statistical significance was reported separately and was not required for this directional criterion.

GDD and viticulture-oriented thresholds. Annual GDD summed the positive daily excess of Tmean above a 10 °C base on identical daily support. Requiring at least five valid growing seasons per station retained 144 stations in Spain, 36 in France, 10 in Greece, and 261 in Italy for the primary GDD error analysis. Classical boundaries at 1390, 1670, 1940, and 2220 degree-days classified each station-year GDD value into one of five Winkler regions. For each station, same-region agreement equaled the proportion of matched years assigned to the same region as the observed reference; country summaries then averaged these station-level proportions with equal weight. April–May frost days (TN < 0 °C) and June–August hot days (TX ≥ 35 °C) entered the analysis only when every constituent month reached at least 80% completeness. These thresholds are operational sensitivity indicators rather than observations of cultivar-specific phenology.

Known provenance limits. The analytical archive did not retain exact E-OBS and MARS-STAT/JRC Agri4Cast construction-station membership; the study therefore evaluated these datasets against the available station reference rather than claiming fully independent validation. AgERA5 supplied local-time daily statistics. ERA5 and ERA5-Land originated from hourly fields, and their daily extrema can depend on whether aggregation uses Coordinated Universal Time (UTC) or a local-time shift. The archive retained provider date labels but not hourly timestamps, retrieval-time shifts, or station observing-day conventions. Same-date pairing therefore did not establish an identical 24-hour window across datasets.

Reproducibility. All reported bootstrap intervals and rank probabilities used fixed seeds derived from SHA-256 labels. Daily, anomaly, GDD, and viticulture-threshold paired rankings used 5,000 station-bootstrap replicates; the primary 10-year and five-year seasonal analyses used 2,000 and 1,000 replicates, respectively. Deterministic scripts maintained by the author generated all numerical results and scientific figures. ChatGPT assisted with text revision and the drafting and review of analytical and document-processing scripts; the author independently reviewed the scripts and verified the resulting statistics and figures.

## Supplementary Figures

Figures S1–S18 provide the additional station-scale, seasonal, robustness, anomaly, and viticulture-oriented diagnostics cited in the main manuscript. No separate data tables, source datasets, or code files accompany this submission.

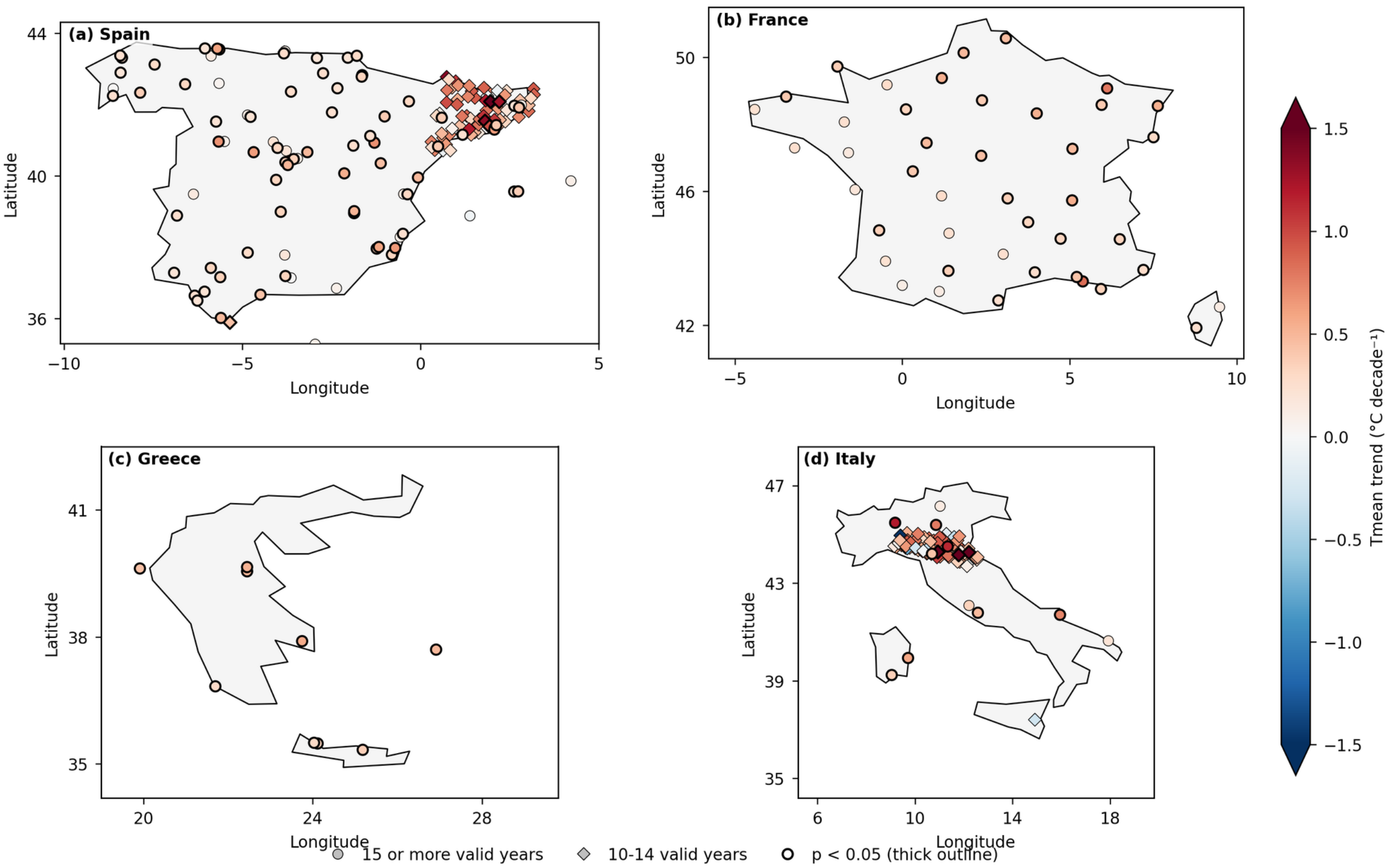

***Figure S1. Station-specific observed annual Tmean trends for Spain, France, Greece, and Italy. Greek observations are from HNMS/EMY; the other countries use ECA&D. The maps show slopes only for records with at least 10 valid annual values. Thick outlines denote unadjusted station-specific OLS p < 0.05; the maps are descriptive and do not apply HAC, FDR, or station-bootstrap criteria.***

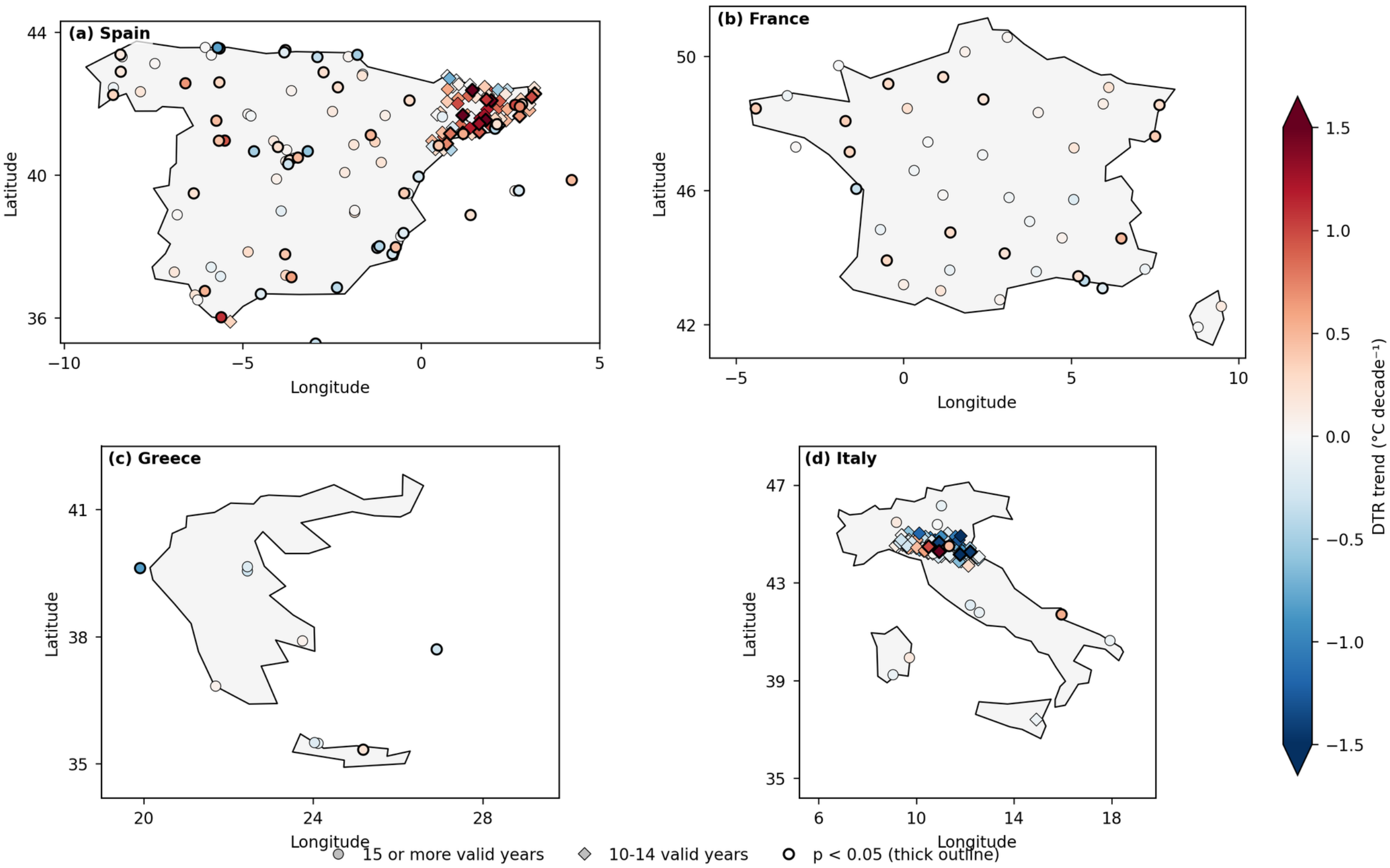

***Figure S2.*** *Station-specific observed annual DTR trends. Record windows differ among stations; the maps therefore describe local heterogeneity rather than a common-period national trend. Thick outlines denote unadjusted station-specific OLS $p < 0.05$; the maps are descriptive and do not apply HAC, FDR, or station-bootstrap criteria.*

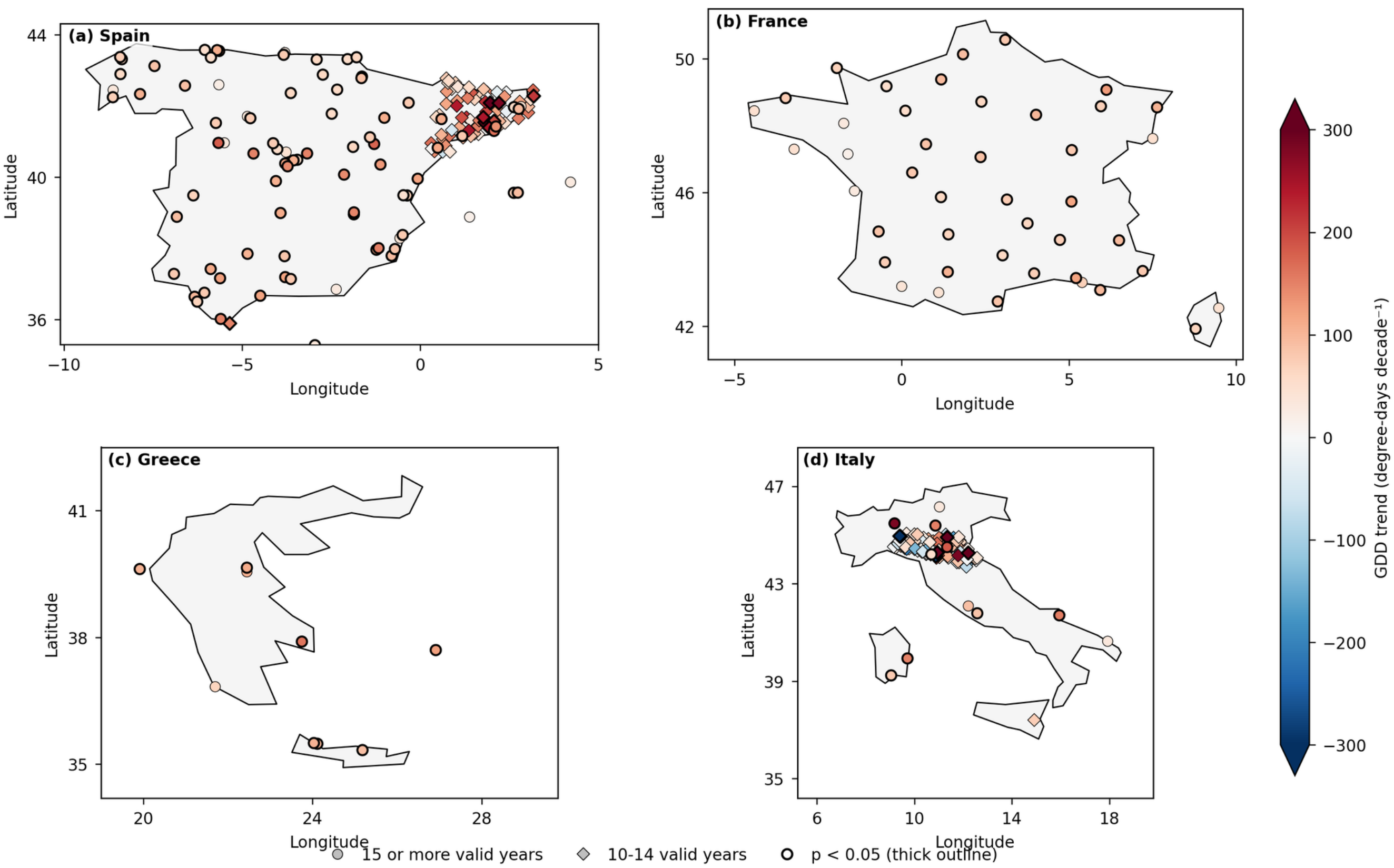

***Figure S3. Station-specific observed April–October GDD trends. The figure shows positive and negative slopes together with record lengths; the main manuscript reports the robustness of the country-level GDD trends. Thick outlines denote unadjusted station-specific OLS $p < 0.05$; the maps are descriptive and do not apply HAC, FDR, or station-bootstrap criteria.***

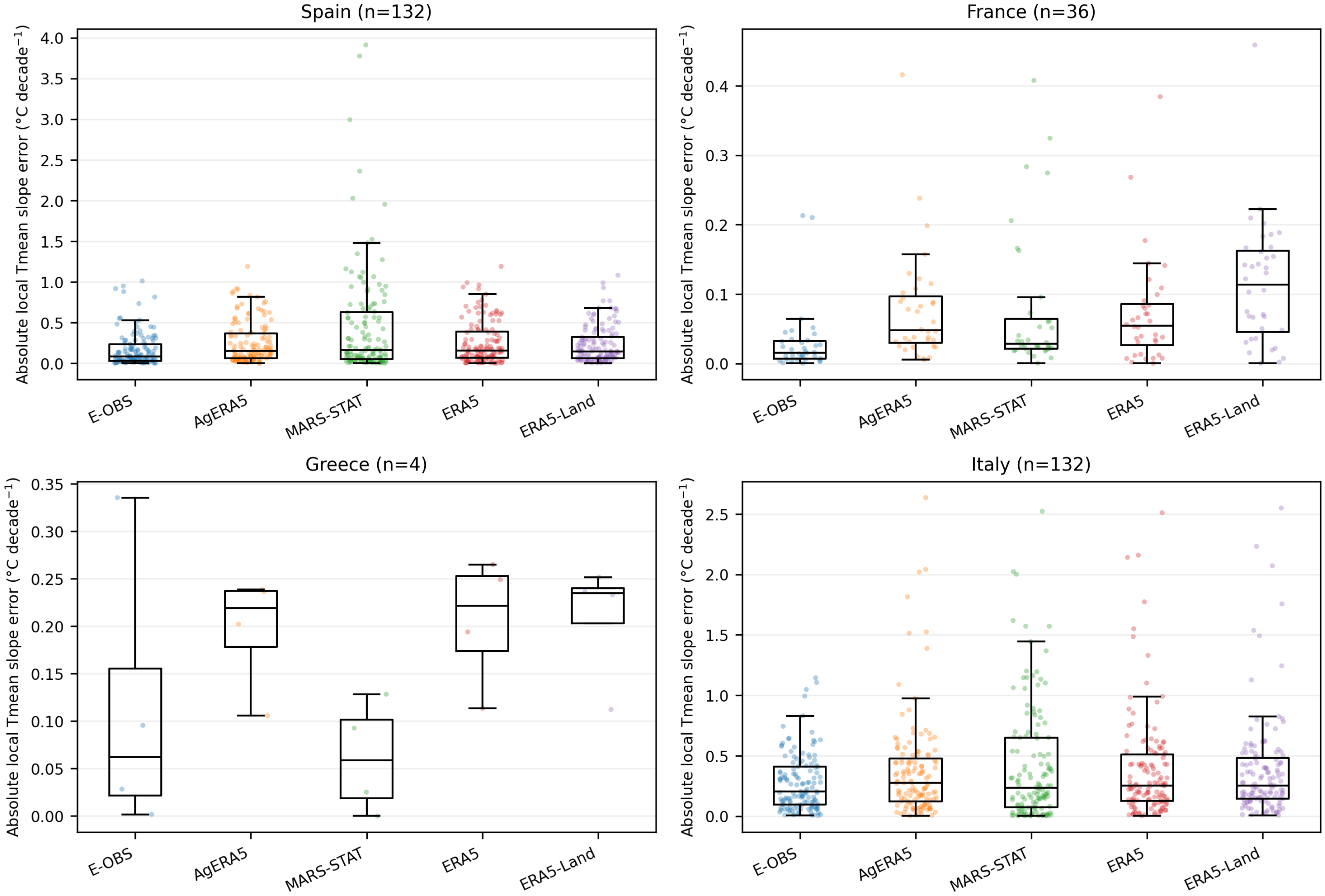

***Figure S4. Dataset-minus-observed local Tmean trend errors on the identical station intersection available for all five gridded datasets. The panels display all retained station observations; panel titles report common-support station counts. MARS-STAT denotes MARS-STAT/JRC Agri4Cast.***

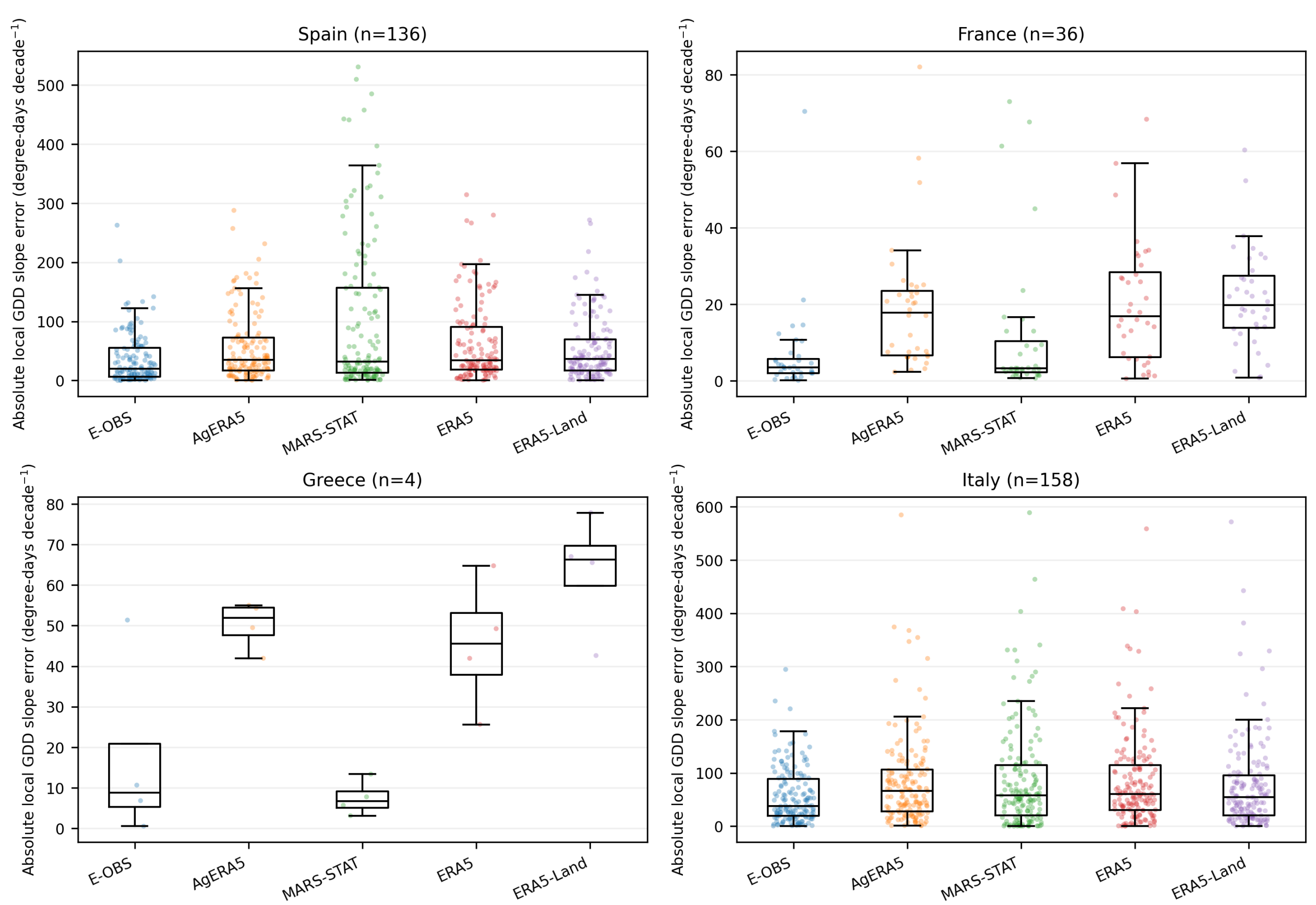

***Figure S5. Dataset-minus-observed local GDD trend errors on the identical station intersection available for all five gridded datasets. The panels display all retained station observations; panel titles report common-support station counts. MARS-STAT denotes MARS-STAT/JRC Agri4Cast.***

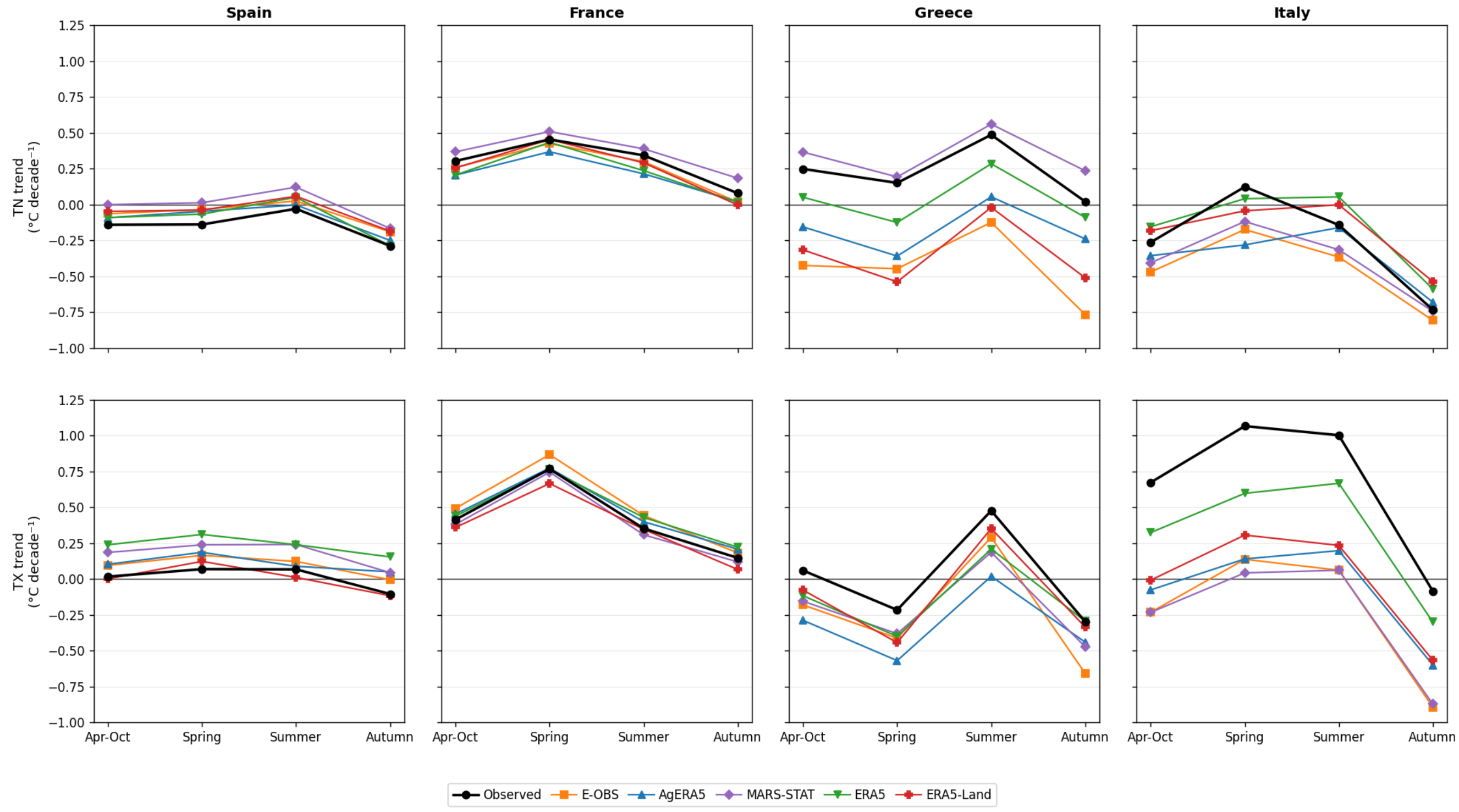

***Figure S6. Country-balanced observed and gridded TN and TX trends for the April–October growing season, spring, summer, and autumn. These are OLS point estimates. The main manuscript summarizes seasonal HAC, station-bootstrap, fixed-network, and false discovery rate results; Figure S13 is limited to annual robustness diagnostics. MARS-STAT denotes MARS-STAT/JRC Agri4Cast.***

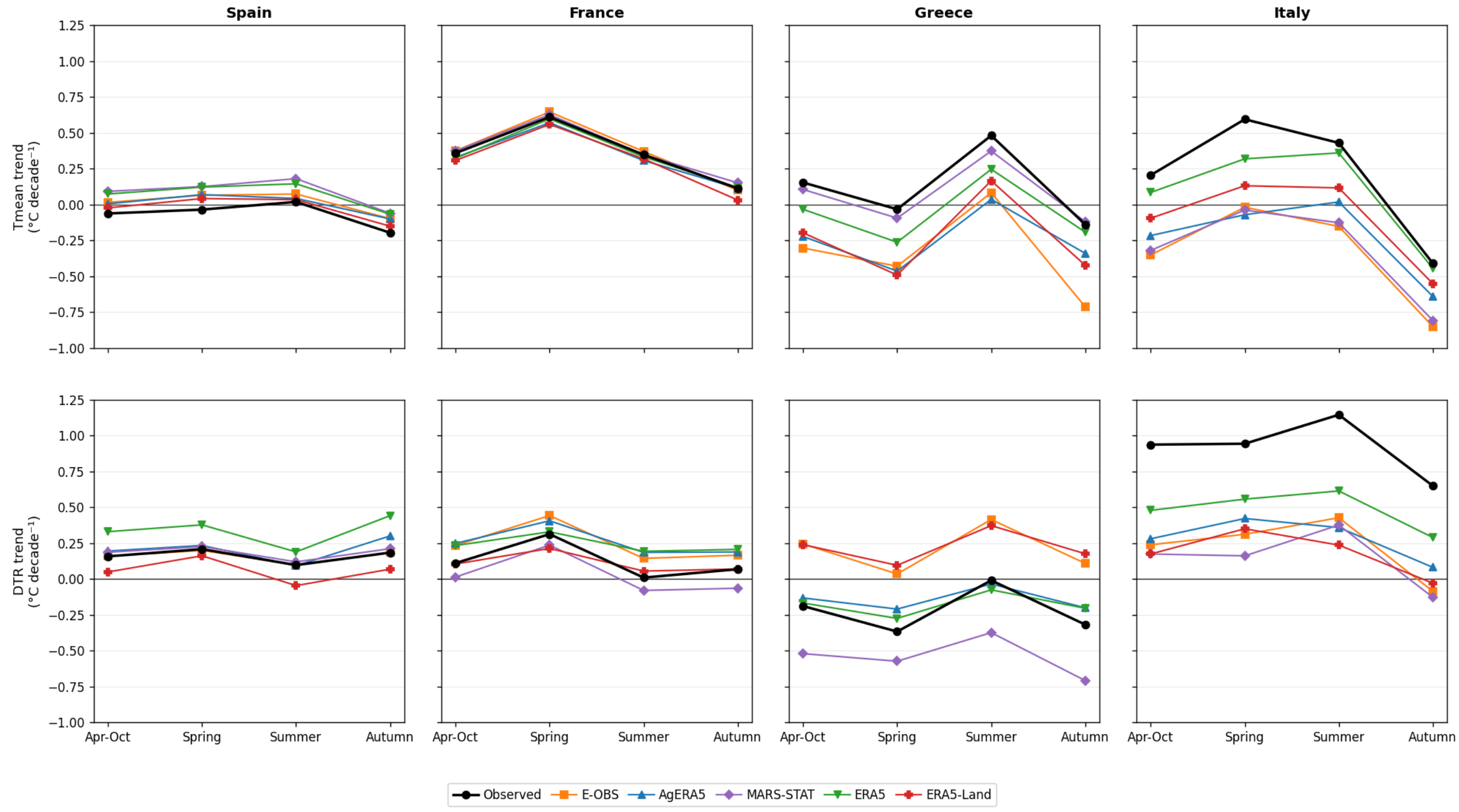

***Figure S7. Country-balanced observed and gridded Tmean and DTR trends for the April–October growing season, spring, summer, and autumn. These calendar subdivisions are thermal periods, not measured phenological stages. The main manuscript summarizes seasonal robustness; Figure S13 is limited to annual diagnostics. MARS-STAT denotes MARS-STAT/JRC Agri4Cast.***

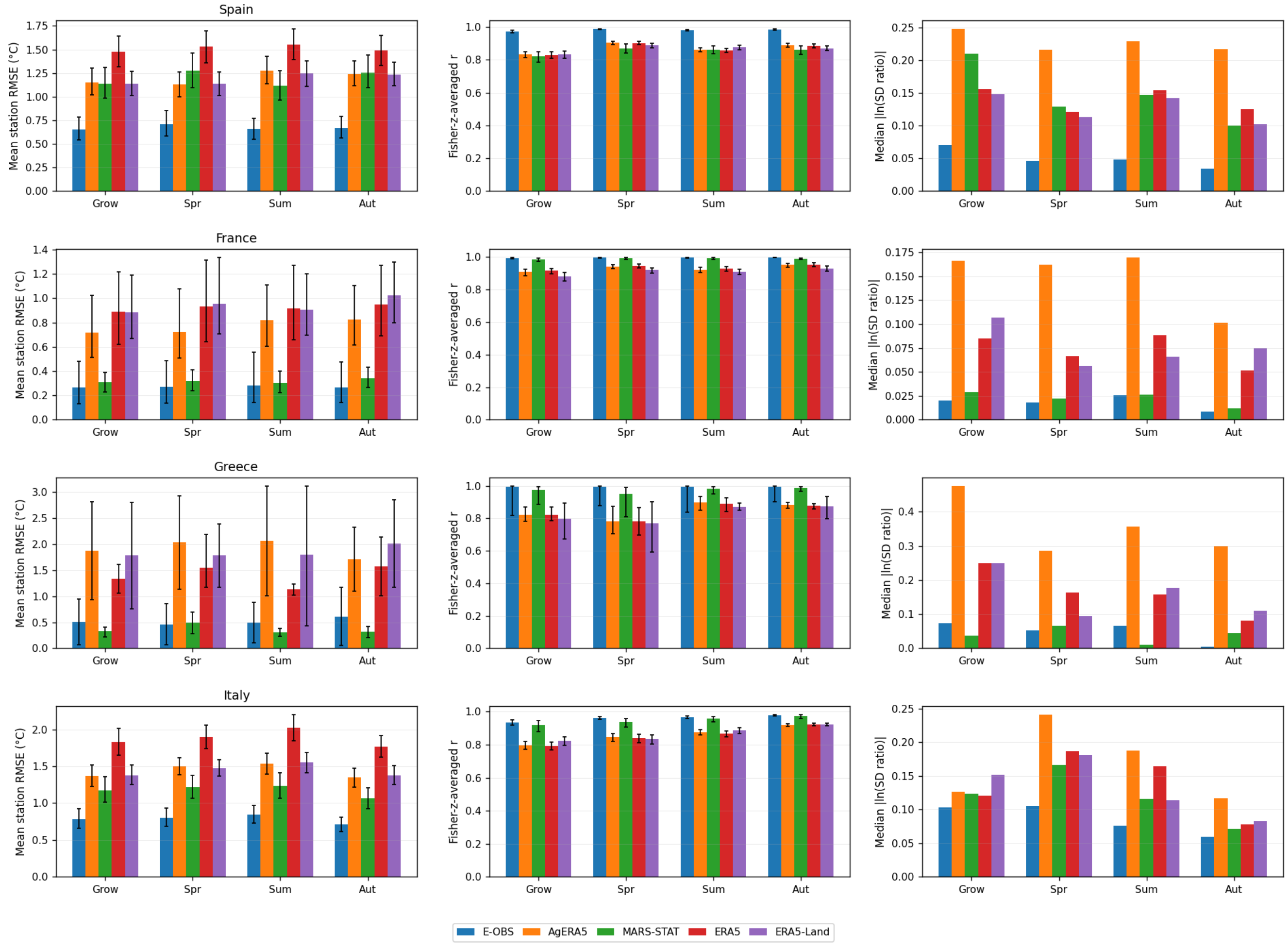

***Figure S8. Primary common-support seasonal validation for daily minimum temperature (TN). Rows are countries; columns show mean station-level root mean square error (RMSE), Fisher-z-averaged r, and the median absolute natural log-ratio of the dataset-to-observed standard deviation [median |ln(SD ratio)|] for April–October (Grow), spring (April–May; Spr), summer (June–August; Sum), and autumn (September–October; Aut). The seasonal analysis transformed station correlations with Fisher's z, averaged the transformed values with equal station weight, and back-transformed the mean to r. Only stations with at least 10 matched years contribute. Error bars show paired station-bootstrap 95% intervals for RMSE and r; the variability column shows point estimates, with 0 indicating exact variability agreement. MARS-STAT denotes MARS-STAT/JRC Agri4Cast.***

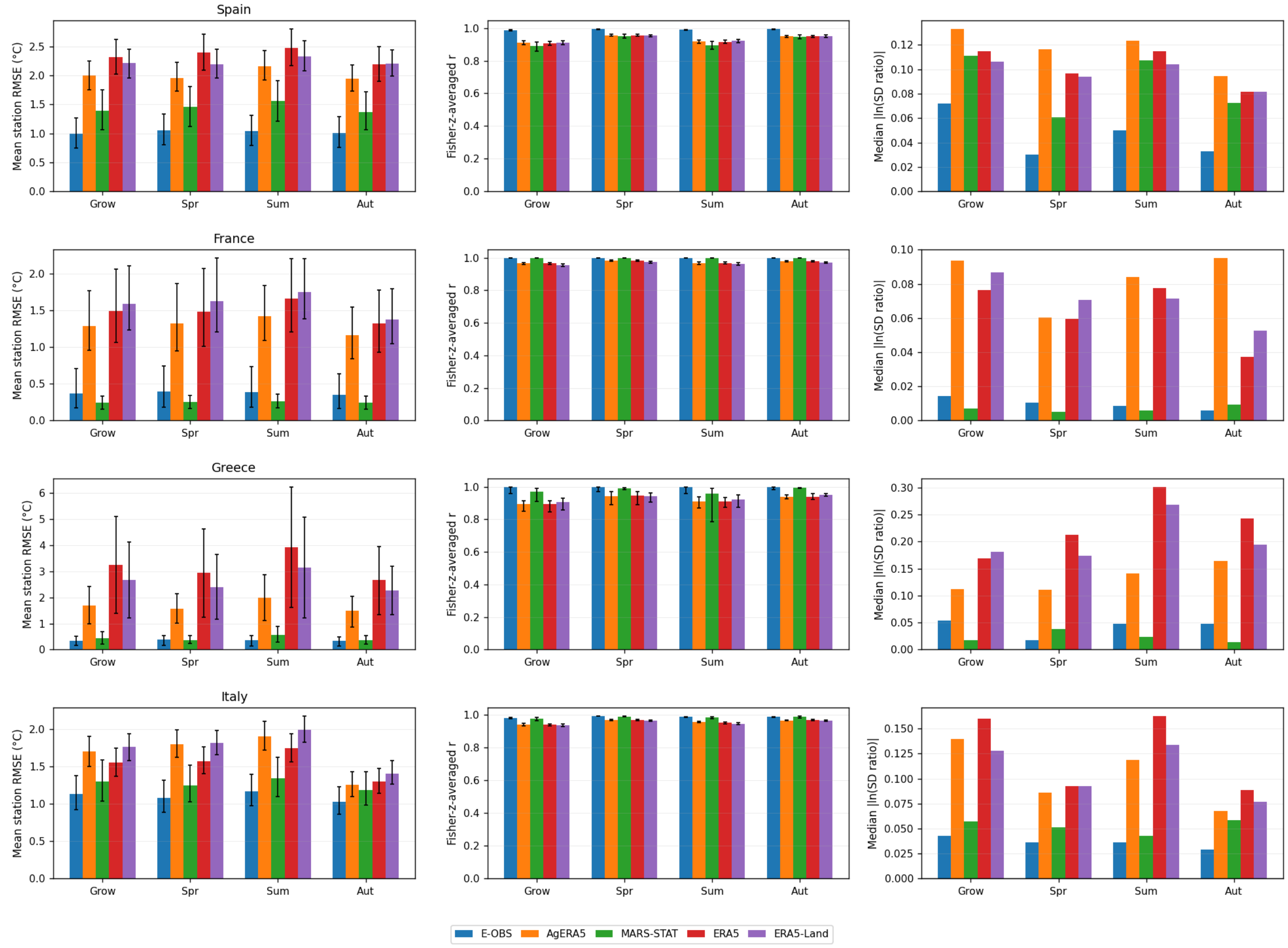

***Figure S9. Primary common-support seasonal validation for daily maximum temperature (TX). Rows are countries; columns show mean station-level root mean square error (RMSE), Fisher-z-averaged r, and the median absolute natural log-ratio of the dataset-to-observed standard deviation [median |ln(SD ratio)|] for April–October (Grow), spring (April–May; Spr), summer (June–August; Sum), and autumn (September–October; Aut). The seasonal analysis transformed station correlations with Fisher's z, averaged the transformed values with equal station weight, and back-transformed the mean to r. Only stations with at least 10 matched years contribute. Error bars show paired station-bootstrap 95% intervals for RMSE and r; the variability column shows point estimates, with 0 indicating exact variability agreement. MARS-STAT denotes MARS-STAT/JRC Agri4Cast.***

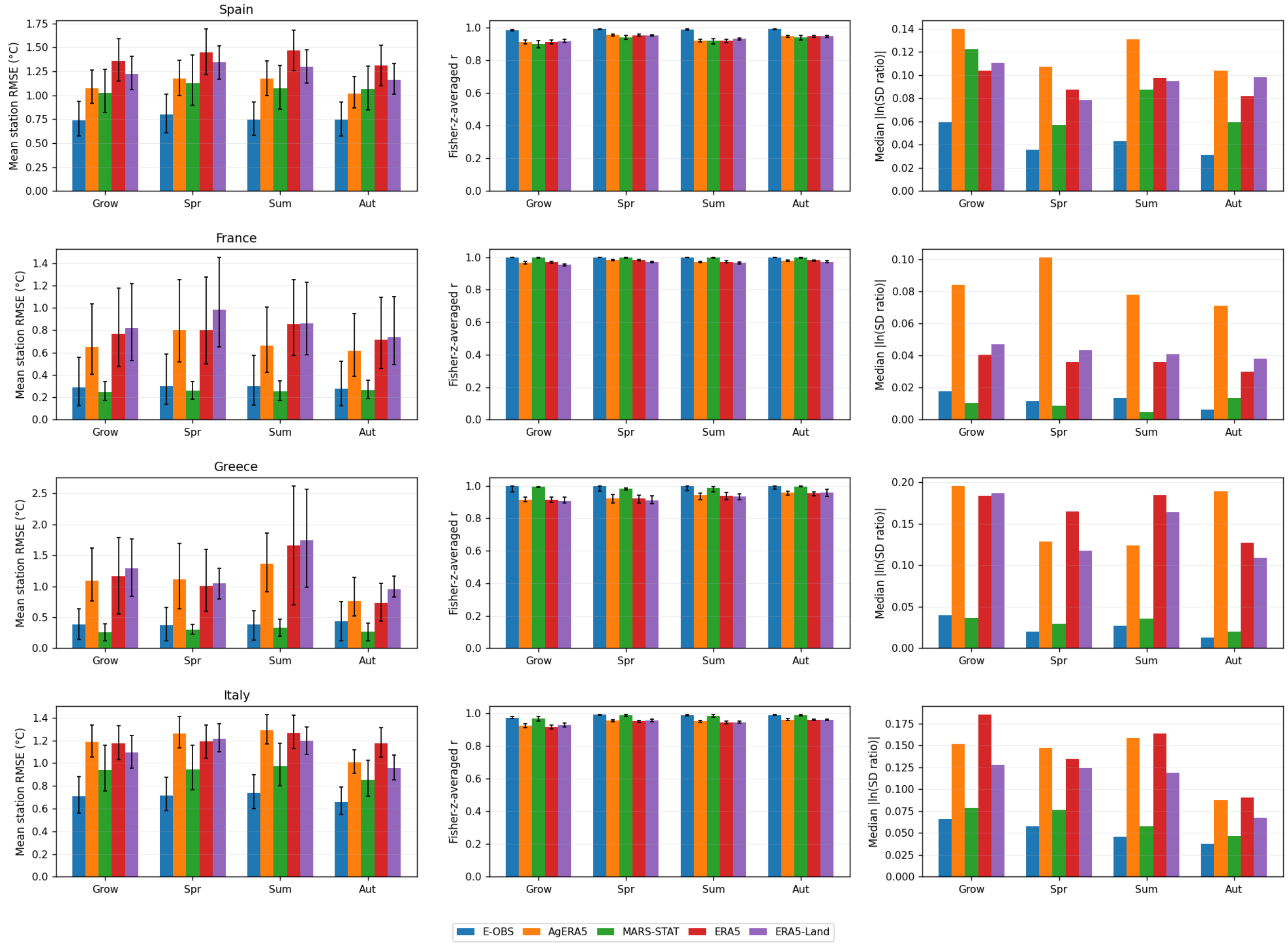

***Figure S10. Primary common-support seasonal validation for mean temperature (Tmean). Rows are countries; columns show mean station-level root mean square error (RMSE), Fisher-z-averaged r, and the median absolute natural log-ratio of the dataset-to-observed standard deviation [median |ln(SD ratio)|] for April–October (Grow), spring (April–May; Spr), summer (June–August; Sum), and autumn (September–October; Aut). The seasonal analysis transformed station correlations with Fisher's z, averaged the transformed values with equal station weight, and back-transformed the mean to r. Only stations with at least 10 matched years contribute. Error bars show paired station-bootstrap 95% intervals for RMSE and r; the variability column shows point estimates, with 0 indicating exact variability agreement. MARS-STAT denotes MARS-STAT/JRC Agri4Cast.***

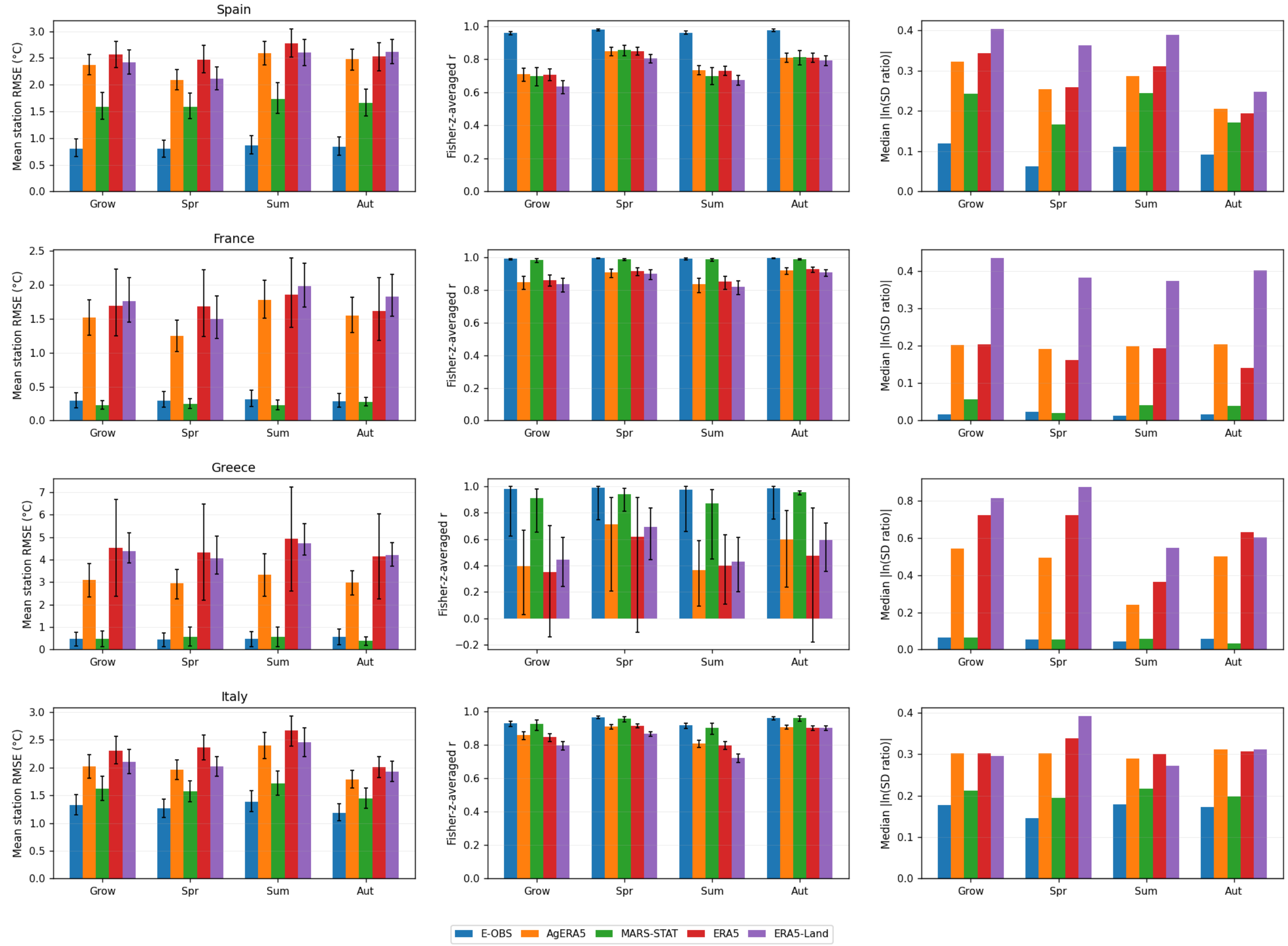

***Figure S11. Primary common-support seasonal validation for diurnal temperature range (DTR). Rows are countries; columns show mean station-level root mean square error (RMSE), Fisher-z-averaged r, and the median absolute natural log-ratio of the dataset-to-observed standard deviation [median |ln(SD ratio)|] for April–October (Grow), spring (April–May; Spr), summer (June–August; Sum), and autumn (September–October; Aut). The seasonal analysis transformed station correlations with Fisher's z, averaged the transformed values with equal station weight, and back-transformed the mean to r. Only stations with at least 10 matched years contribute. Error bars show paired station-bootstrap 95% intervals for RMSE and r; the variability column shows point estimates, with 0 indicating exact variability agreement. MARS-STAT denotes MARS-STAT/JRC Agri4Cast.***

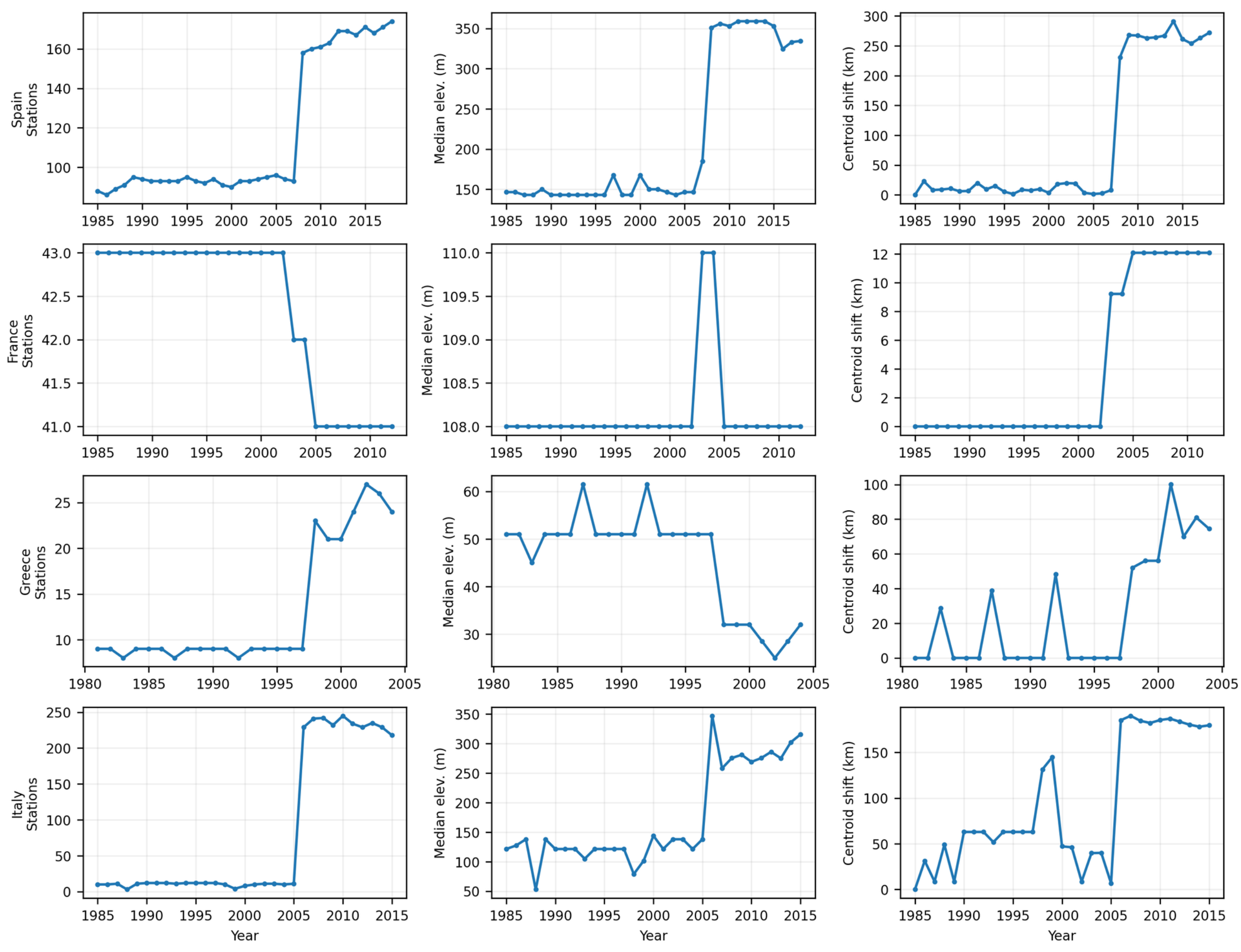

***Figure S12.*** *Annual observed-network composition by country, including station count, median elevation, and spatial centroid. Centroid shift is measured relative to each country's first available annual-network centroid. Changes in network composition motivate the fixed-network and spatial-weighting sensitivities.*

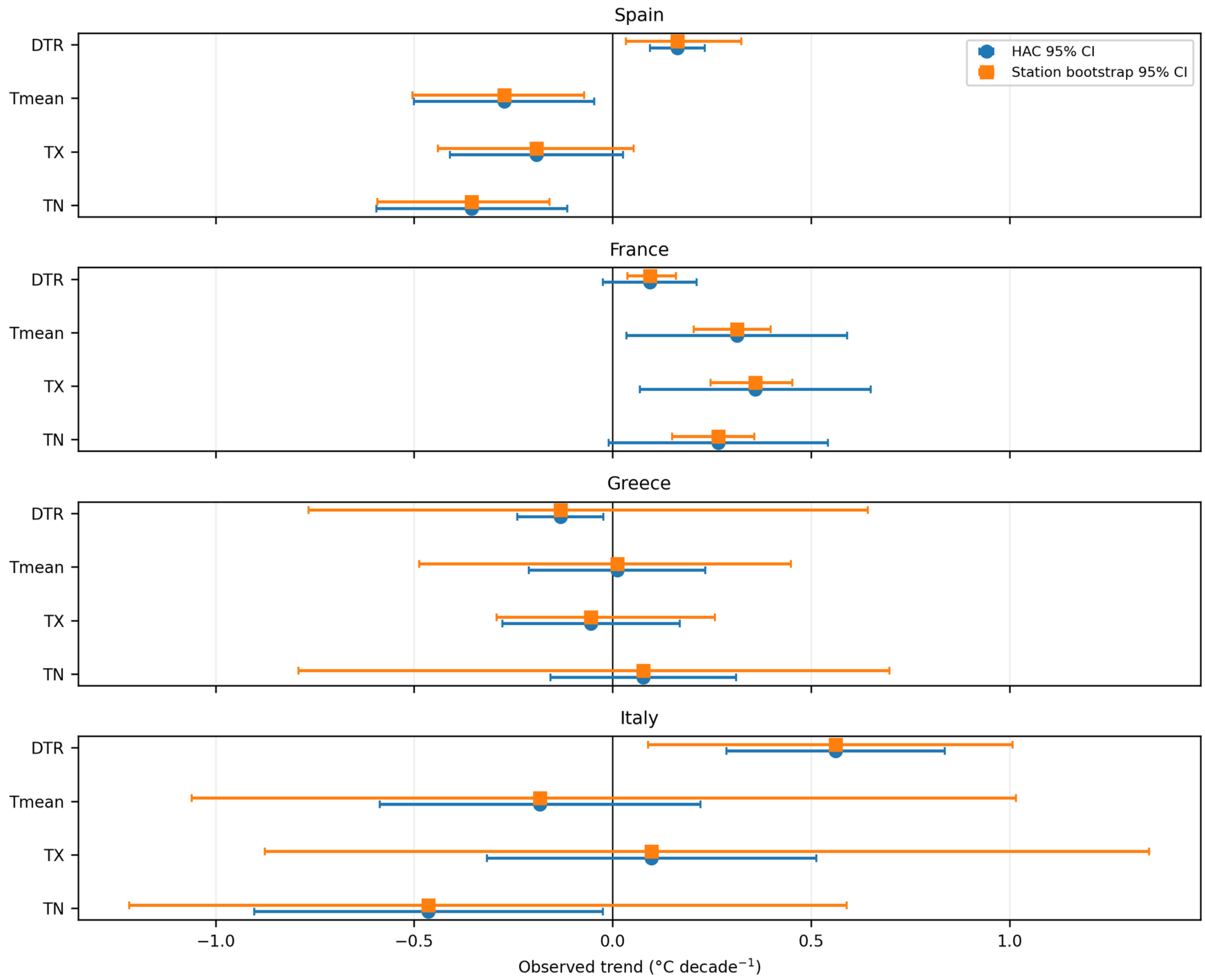

***Figure S13. Observed annual trend robustness for TN, TX, Tmean, and DTR. The figure pairs point estimates with 95% Newey–West heteroskedasticity- and autocorrelation-consistent (HAC) intervals and 95% station-bootstrap intervals; the main manuscript summarizes fixed-network, false discovery rate, and spatial-weighting results.***

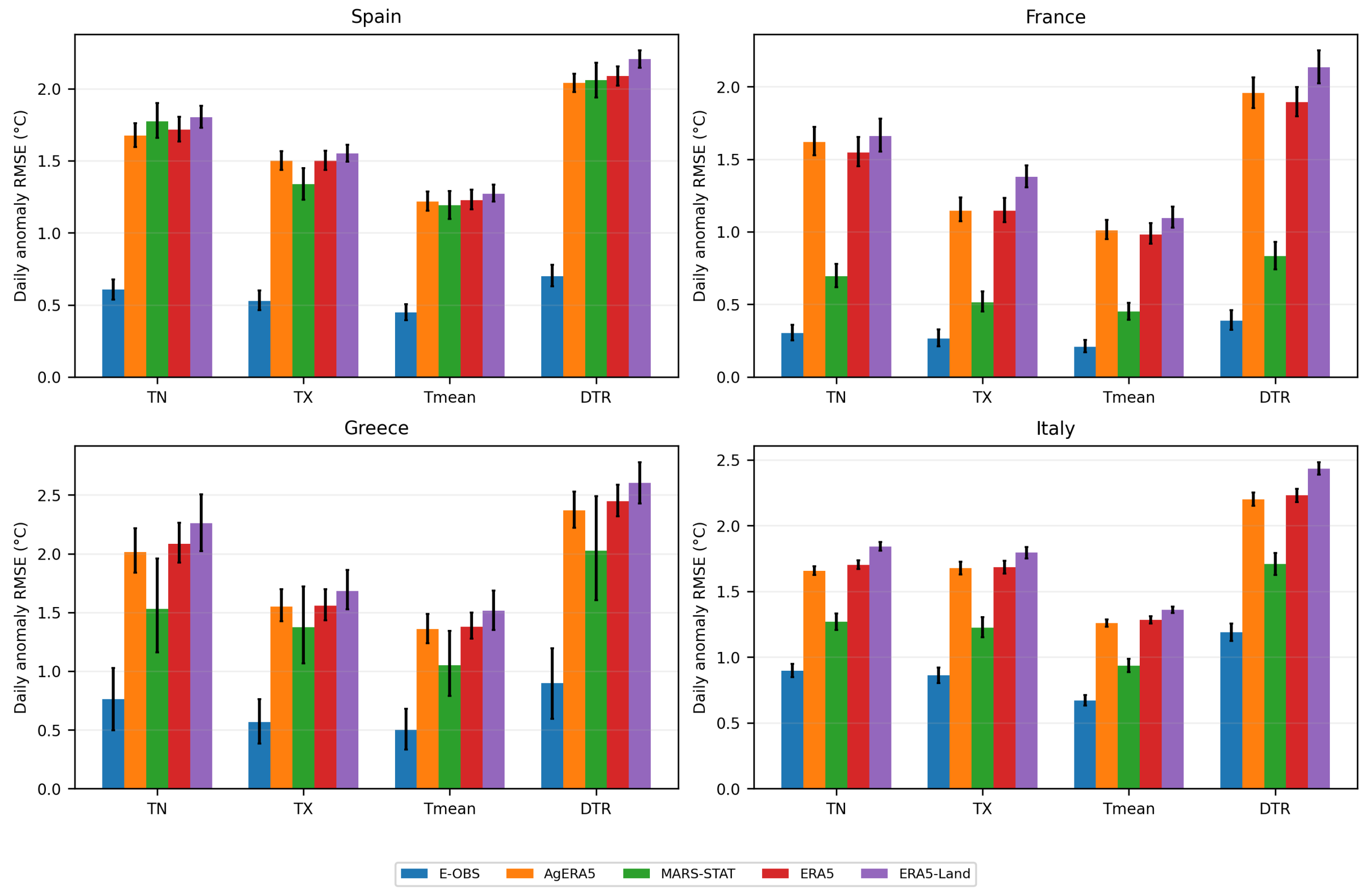

***Figure S14. Common-support monthly-climatology-removed daily-anomaly performance for TN, TX, Tmean, and DTR. For each station and dataset, the procedure removed monthly climatology before calculating station-level metrics. MARS-STAT denotes MARS-STAT/JRC Agri4Cast.***

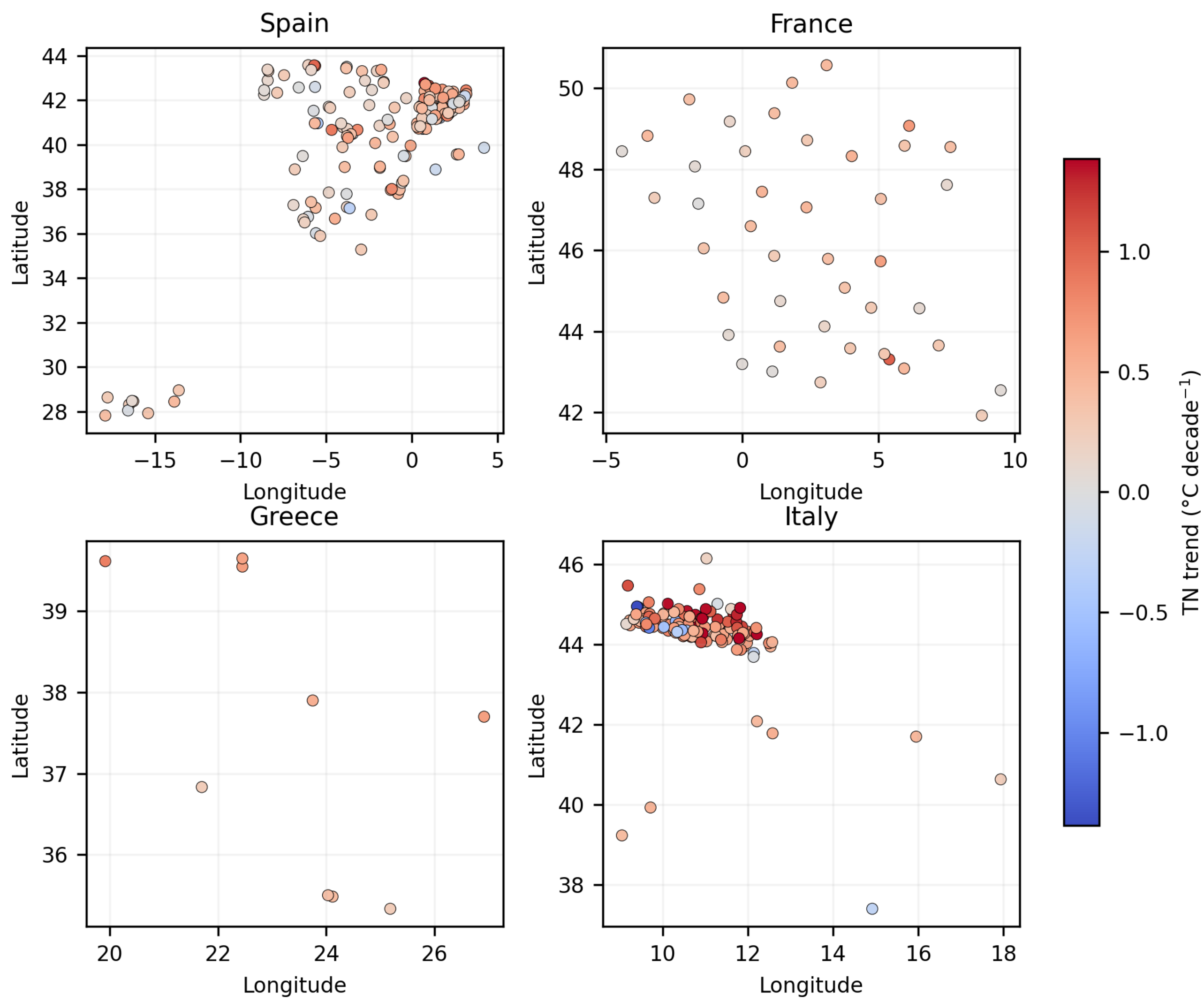

***Figure S15. Station-specific observed annual TN trends. The maps show only stations with at least 10 valid annual values. The figure presents local TN behavior separately from Tmean because trends in daily minima and maxima can differ.***

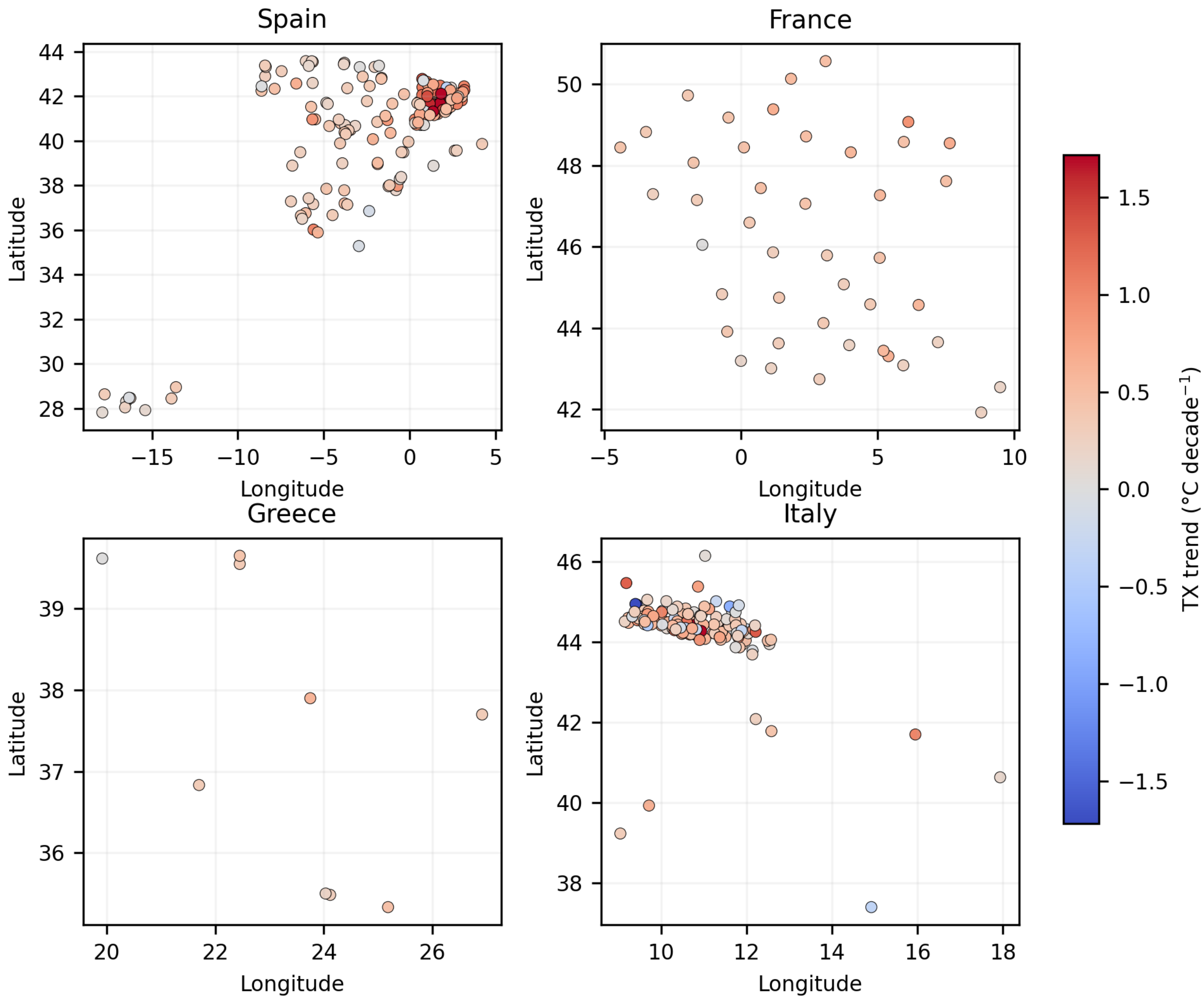

***Figure S16. Station-specific observed annual TX trends. The maps show only stations with at least 10 valid annual values. The figure presents local TX behavior separately from Tmean because trends in daily maxima and minima can differ.***

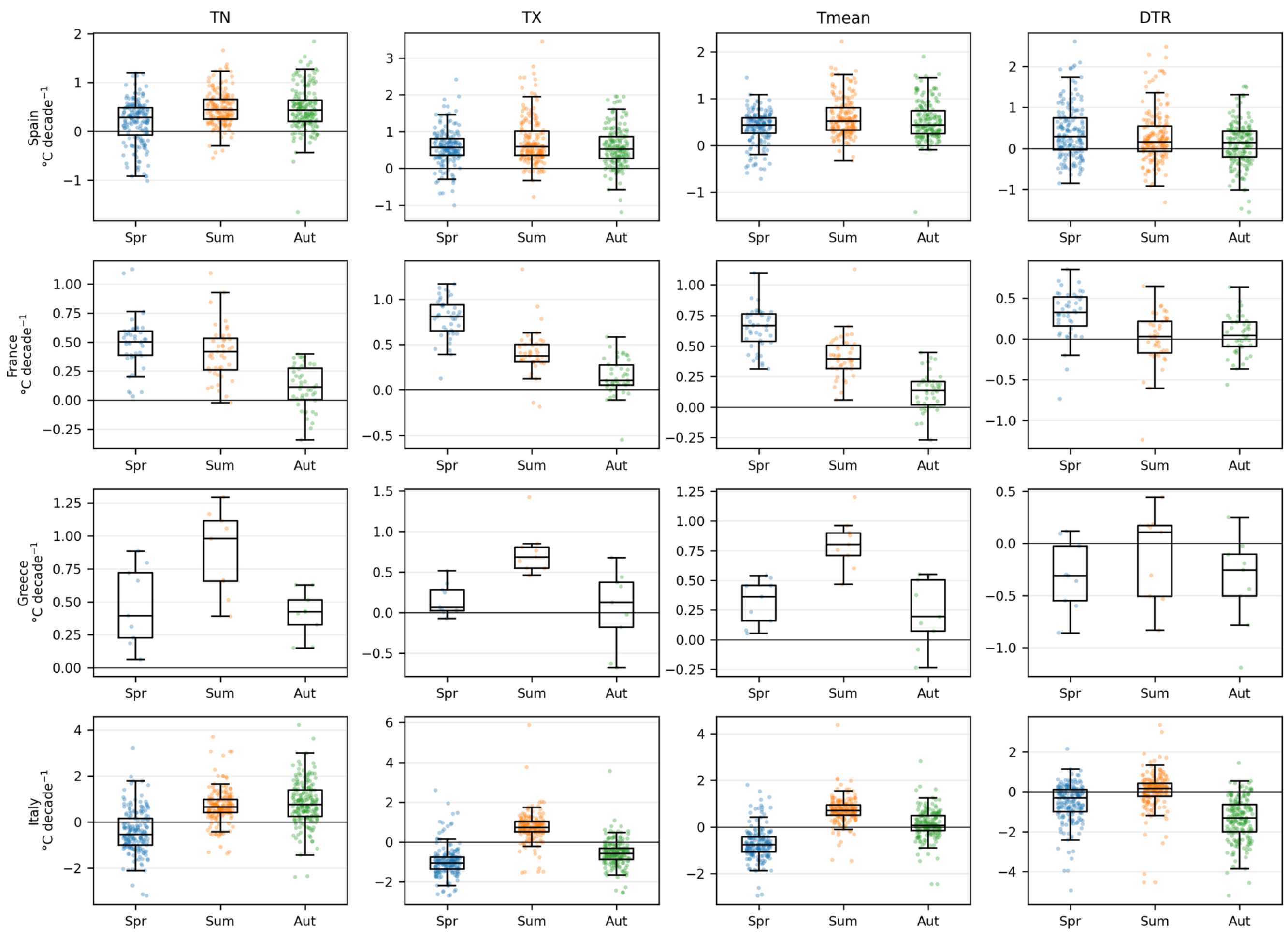

***Figure S17. Distributions of station-specific observed spring, summer, and autumn TN, TX, Tmean, and DTR slopes by country. Only station-period series with at least 10 valid years contribute. These distributions reveal network heterogeneity that is not visible in a single country-balanced slope.***

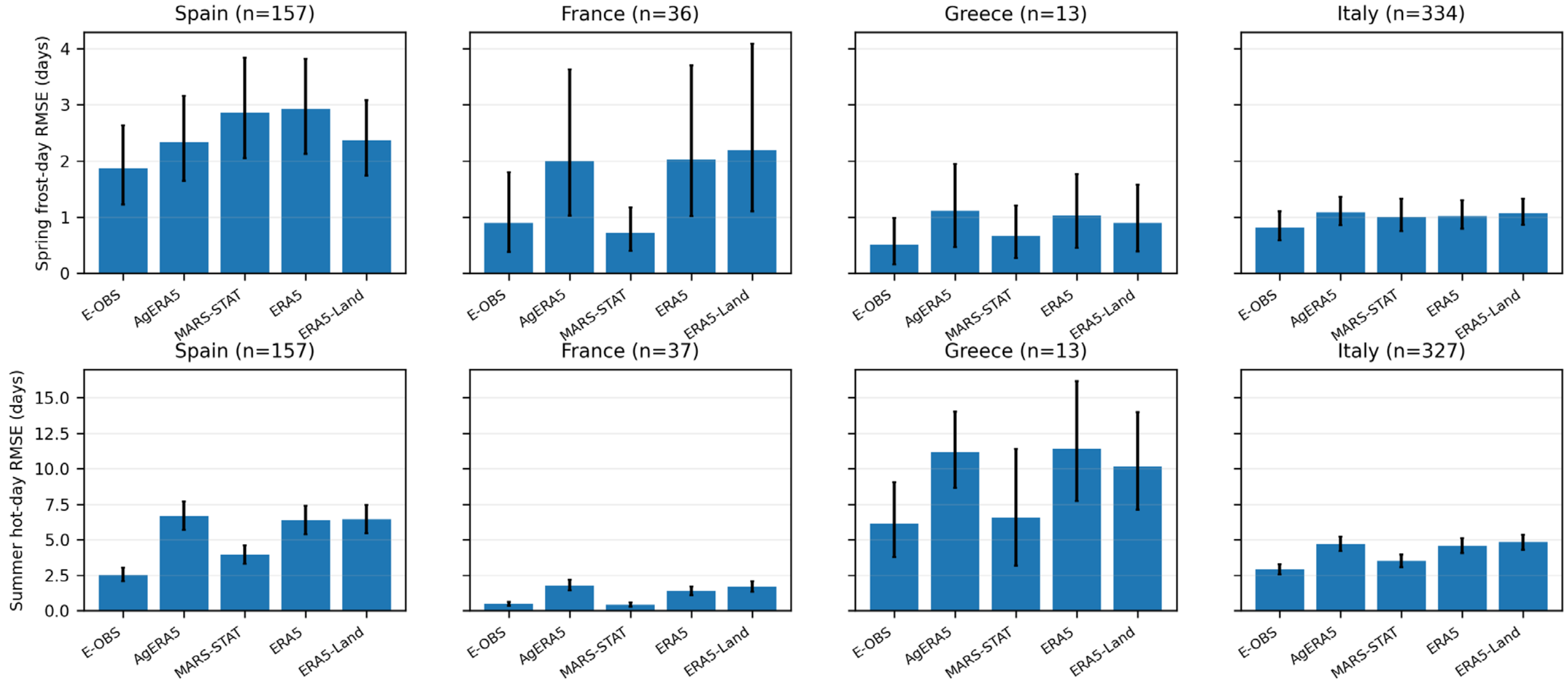

***Figure S18. Viticulture-oriented threshold sensitivity on identical support: April–May frost days (TN < 0 °C) and June–August hot days (TX ≥ 35 °C). Every constituent month must reach at least 80% completeness; error bars show paired station-bootstrap 95% intervals. The thresholds are operational diagnostics rather than cultivar-specific phenological endpoints. MARS-STAT denotes MARS-STAT/JRC Agri4Cast.***